\documentclass[twocolumn]{aastex631}

\usepackage{longtable}
\usepackage{float}

\newcommand{\msun}{\mbox{$M_\odot$}}

\shorttitle{Terzan 5 X-ray sources analysis}
\shortauthors{G. Kumawat et al.}
\begin{document}

\title{A Comprehensive Analysis of X-ray Sources in Terzan 5 Using Chandra Observations}

\correspondingauthor{Gourav Kumawat} \email{kumawat1@ualberta.ca}

\author[0009-0001-2894-0155]{Gourav Kumawat}
\affiliation{Department of Physics, University of Alberta, Edmonton AB T6G 2G7, Canada}

\author[0000-0003-3944-6109]{Craig O. Heinke}
\affiliation{Department of Physics, University of Alberta, Edmonton AB T6G 2G7, Canada}

\author[0000-0002-7716-1166]{Jiaqi Zhao}
\affiliation{Department of Physics, University of Alberta, Edmonton AB T6G 2G7, Canada}

\author[0000-0003-2506-6041]{Arash Bahramian}
\affiliation{International Centre for Radio Astronomy Research, Curtin University, Kent St, Bentley WA 6102, Australia}

\author[0000-0002-9673-7802]{Haldan N. Cohn}
\affiliation{Department of Astronomy, Indiana University, 727 E. Third St., Bloomington, IN 47405, USA}

\author[0000-0001-9953-6291]{Phyllis M. Lugger}
\affiliation{Department of Astronomy, Indiana University, 727 E. Third St., Bloomington, IN 47405, USA}



\begin{abstract}

We analyze photometry, spectra, and variability of over 100 faint X-ray sources in the globular cluster Terzan 5, using 737 ks of \textit{Chandra} data. X-ray colors and spectral fitting allow clear separation of foreground sources (with less extinction than the cluster), quiescent low-mass X-ray binaries (qLMXBs), and sources with harder spectra. We identify 22 candidate qLMXBs, over twice that found in any other cluster. This is consistent with Terzan 5's stellar interaction rate, the highest among Galactic globular clusters. We do not see qLMXBs dominated by thermal emission below $L_X\sim10^{32}$ erg/s, though qLMXBs with stronger nonthermal emission could be missed. We find that more than 50\% of the qLMXB sources have neutron star thermal component contributing over 80\% of the total luminosity. We  report an unusual spectral feature around 1.75~keV in the combined spectrum of Ter~5~X-3. The concentration of the qLMXBs within the cluster is consistent with that of a population of mass $1.46 \pm 0.14$~M$_\odot$. We identify secure X-ray counterparts to millisecond pulsars Terzan 5 ar and Terzan 5 at, using positional coincidence and orbital X-ray light curves matching those expected for spider pulsars. 

\end{abstract}



\section{Introduction} \label{sec:intro}

Star clusters offer critical insights into the workings of stellar evolution and dynamics, functioning as natural testbeds for astrophysical theories. Among these, globular clusters (GCs) are particularly important due to their ancient origin and exceptionally dense stellar environments. With its sub-arcsecond angular resolution, the \textit{Chandra X-ray Observatory} has made it possible to resolve both bright and faint X-ray sources, even in the densely populated cores of clusters \citep{Grindlay01, VerbuntLewin2006}. Owing to the dynamical processes in such environments, GCs make ideal settings for the frequent formation of exotic stellar systems—including low-mass X-ray binaries (LMXBs), millisecond pulsars (MSPs), and cataclysmic variables (CVs)—which are observed in much greater numbers in GCs compared to the Galactic field \citep{Clark1975,  Pooley03, Heinke2003}.

These exotic systems are thought to arise predominantly through binary evolutionary pathways, often influenced or accelerated by dynamical interactions such as exchange encounters and stellar collisions \citep{Ivanova2006, Ivanova2008}. A notable example is low-mass X-ray binaries (LMXBs)—systems where neutron stars (NSs) accrete material from low-mass stellar companions—which are found to be over 100 times more frequent in GCs per unit stellar mass than in the Galactic field \citep{Clark1975}. This enhancement is attributed to the ability of NSs to acquire new companions through exchange interactions following the disruption of their initial binaries by supernova explosions \citep{Verbunt1987}. Among faint X-ray sources ($L_X < 10^{34.5}$ erg s$^{-1}$), the most luminous on average are quiescent low-mass X-ray binaries (qLMXBs), where the NS accretes at a very low or negligible rate. These systems typically exhibit X-ray luminosities of $10^{32-33}$ erg s$^{-1}$. They often show soft, blackbody-like spectra, generally interpreted as thermal radiation from the NS surface or residual low-level accretion activity, and commonly also include a harder spectral component fit by a power-law \citep{Rutledge2002, Heinke2003}. Their characteristic spectral profiles and X-ray luminosities (significantly more luminous than other objects with soft spectra) facilitate robust identification within globular clusters. The formation of qLMXBs is believed to be closely linked to stellar interactions, with their abundance correlating with the stellar encounter rate of the host cluster \citep{Heinke2003, Pooley2006}.

Millisecond pulsars (MSPs) are commonly regarded as the evolutionary descendants of qLMXBs, where extended episodes of accretion spin up NSs to millisecond rotation periods and simultaneously reduce their magnetic field strengths \citep{Alpar1982, Bhattacharya1991}. This ``recycling" mechanism can span up to a billion years and typically produces MSPs with spin periods shorter than 10 milliseconds, often in binary systems with helium white dwarfs, or with non-degenerate stellar companions, referred to as ``spiders.'' Spider systems are further categorized into two primary types: Black Widows (BWs) and Redbacks (RBs), which have companion masses of $< 0.05\,M_\odot$ and $0.1$--$0.5\,M_\odot$, respectively. These spider systems serve as evolutionary intermediates between LMXBs and canonical MSPs. The discovery of transitional MSPs (tMSPs), which switch between states dominated by accretion and those powered by rotation, has provided crucial observational evidence of this evolutionary pathway \citep{Archibald2009, Papitto2022}. In GCs, most MSPs appear as soft X-ray emitters, with spectra primarily characterized by thermal, blackbody-like emission from heated polar caps on the NS surface \citep{Bogdanov2006}. A subset, however, shows harder X-ray spectra, which can be modelled by a power-law of photon index 1-2. Some, with relatively high spin-down rates (and thus inferred magnetic fields), show sharp X-ray pulsations, indicating synchrotron X-ray emission from particles within the pulsar magnetosphere \citep{Saito97}. The "spider" systems are typically identified by radio eclipses due to absorption or scattering of the pulsar signal in a wind from the non-degenerate companion star. These do not show pulsations, but generally show hard X-ray spectra and often show orbital modulation of the X-rays, due to the complicated interaction of the pulsar wind and the companion star's wind leading to beamed radiation in some directions \citep{Wadiasingh2017}. The dense stellar environments of GCs promote frequent dynamical encounters, making them prolific sites for MSP production, with MSP densities orders of magnitude higher than in the Galactic field \citep{Hui2010, Zhao2022}.

Cataclysmic variables (CVs) are interacting binaries where a white dwarf (WD) accretes mass from a companion star—most often a low-mass main-sequence (MS) star or a subgiant—through Roche lobe overflow. The transferred material generally accumulates into an accretion disk around the WD, where it is heated to X-ray-emitting temperatures as gravitational energy is converted into thermal energy at, or in a shock above, the white dwarf surface. These systems characteristically display hard X-ray spectra due to hot, optically thin plasma emission and possess X-ray luminosities typically below $\sim10^{32}$ erg s$^{-1}$ \citep{Mukai17}. Although globular clusters are effective at dynamically producing CVs, they also tend to disrupt wider binaries that could otherwise evolve into CVs, leading to a complex interplay between creation and destruction processes \citep{Cool2013, Belloni2020}. This dynamic equilibrium makes the origin of the CV population in clusters an active area of research \citep{Cheng2018, Heinke2020}. CVs outnumber neutron star binaries in clusters due to the higher abundance of WDs \citep{Maccarone2007}. In some cases, WDs may accrete from more evolved donors such as red giants via wind accretion, leading to the formation of symbiotic binaries. However, no  symbiotic systems have yet been confirmed in globular clusters \citep{Henleywillis2018, Bahramian2020, Kumawat2024}; symbiotics are  likely rare due to the disruptive effect of dense environments on the wide binaries needed to form such systems \citep{Belloni2020}.

Active binaries (ABs) are close binary systems composed of two stars—generally main-sequence (MS), subgiant, or giant stars—that exhibit high levels of coronal X-ray emission due to rapid stellar rotation and strong magnetic activity \citep{Gudel2004}. This enhanced activity is typically driven by tidal locking in close binary systems, which strengthens dynamo mechanisms and leads to the formation of hot, X-ray-emitting coronal plasma. ABs are commonly found as faint X-ray sources with X-ray luminosities of $\lesssim 10^{31}$ erg s$^{-1}$ and typically exhibit soft X-ray spectra \citep{Dempsey97,Lugger2023}. However, they are challenging to identify in GCs without deep optical data, due to their similar X-ray properties (luminosities, spectra, variability) as CVs \citep{Heinke2005}.

Located in the Galactic Bulge at a distance of $\sim$6.9 kpc, Terzan 5 is heavily obscured ($E(B-V) = 2.28$ mag) and unusually metal-rich for a GC, with [Fe/H] = $-0.23$ \citep{Harris1996}. Terzan 5 is both massive and dense \citep{Lanzoni2010}, giving it the highest stellar encounter rate among all Galactic globular clusters  
(6.8 times that of 47 Tuc),  16\% of the total computed stellar encounter rate of all Galactic globular clusters \citep{Bahramian2013}.

\citet{Heinke2006} utilized early \textit{Chandra} observations to detect 50 X-ray sources down to a limiting luminosity of $3\times10^{31}$ erg s$^{-1}$ within the half-mass radius of Terzan 5, identifying 13 likely quiescent LMXBs based on their X-ray spectra and luminosities. Since then, Terzan 5 has been repeatedly observed with \textit{Chandra}, typically to study the cooling of individual quiescent LMXBs (see below), culminating in over 700 ks of archival \textit{Chandra} data. \cite{Cheng2019} used this deep \textit{Chandra} data to catalog 489 X-ray sources within 4.3 arcmin of the cluster center and identify mass segregation signatures in their radial distribution. \citet{Bahramian2020} presented a \textit{Chandra}/ACIS catalog, including photometry and automated spectral fitting, of over 1,100 faint X-ray sources in 38 globular clusters, including Terzan 5.

There have been several detailed studies of transient LMXBs in Terzan 5. Bursts from an LMXB, likely a transient in outburst, were detected from Terzan 5 by \textit{Hakucho} in 1980 \citep{Makishima1981}; non-burst emission was seen by {\it EXOSAT} in 1984 \citep{Warwick88}. \textit{Chandra} observed Terzan 5 during a bright LMXB outburst in 2000 \citep{Heinke2003exo}, producing a sub-arcsecond position for the active LMXB which has since been referred to as EXO 1745-248 (though it's not certain the name is accurate, as a different LMXB might have been seen by EXOSAT). Subsequent outbursts of this LMXB, EXO 1745-248, were observed in 2011 \citep{Altamirano2012} and 2015 \citep{Tetarenko2016}. Studies of EXO 1745-248 in quiescence have shown an 
unusually hard and strongly variable quiescent X-ray spectrum, suggesting contributions from a pulsar wind or residual accretion down to the magnetospheric radius \citep{Wijnands2005,  RiveraSandoval2018}. The precise position of EXO 1745-248 has been measured with the VLA by \citet{Tetarenko2016}. 

IGR J17480-2446 (or Ter 5 X-2) is a transient accreting millisecond X-ray pulsar (AMXP) discovered during its single 2010 outburst by \textit{INTEGRAL}, with a (relatively slow) pulsation frequency of 11~Hz \citep{Papitto2011}. A rare eclipse of Ter 5 X-2 by Earth's moon allowed \citet{Riggio2012} to derive a subarcsecond position. Post-outburst observations in quiescence with \textit{Chandra} \citep{Degenaar2011,Degenaar2013,Degenaar15} revealed significant cooling of the accretion-heated neutron star crust, offering valuable constraints on the thermal response of neutron star interiors to episodic accretion.

Swift J174805.3-244637 (or Ter 5 X-3) was tracked during its 2012 outburst rise with \textit{Swift}/XRT, showing clear spectral hardening during the rise \citep{Bahramian2014}. Archival \textit{Chandra} observations in quiescence showed variability in the nonthermal component, while the thermal component appeared stable \citep{Bahramian2014}, apart from slight cooling directly after its outburst \citep{Degenaar15}. The inferred thermal luminosity suggested a recurrence timescale of $\sim$10 years, assuming only slow neutrino cooling processes operate in the neutron star core. This recurrence timescale was confirmed by detection of a new outburst from Ter 5 X-3 in 2023 \citep{Heinke2023}. 

Terzan 5 hosts the largest known population of radio MSPs among all Galactic globular clusters, likely due to its exceptionally high stellar density and encounter rate. \citet{Lyne1990} discovered the eclipsing MSP Terzan 5 A, now considered a redback MSP. \citet{Ransom2005} discovered 21 new MSPs in Terzan 5 using the Green Bank Telescope, including the fastest spinning neutron star known, Ter 5 ad at 716~Hz \citep{Hessels2006}. 49 MSPs have now been discovered in Terzan 5  \citet{Lyne2000,Cadelano18,Ridolfi21,Padmanabh2024}. Radio timing of these MSPs has been used to measure the cluster mass and structural parameters \citep{Prager2017}. \citet{Bogdanov2021} identified plausible \textit{Chandra} X-ray counterparts to 8 MSPs in Terzan 5, including the 4 spider pulsars then known, and identified orbital modulations of the X-ray emission for three of the spider pulsars (and large variations for the fourth). 

Interferometric radio imaging of Terzan 5 is another way to probe the radio pulsar population. 
\citet{FruchterGoss1990} and  \citet{FruchterGoss2000} analyzed the integrated radio flux density obtained by imaging several globular clusters, and inferred that Terzan 5 likely contains a substantial population of 60-200 undetected MSPs, more than in any other Galactic cluster. Other calculations, using multiple methods, agree that 
$\sim$100--200 MSPs may reside in the cluster \citep{Bagchi2011, Zhao2022, Yin2024}. \citet{Bahramian2018} identified a radio counterpart to the brightest of the faint X-ray sources, and suggested that its pattern of alternating radio-bright and X-ray-bright states indicated that it may be a transitional MSP. \citet{Urquhart2020} used deep VLA radio continuum observations to identify 24 compact radio sources (likely MSPs) in Terzan 5, including three new spider MSP candidates. One of these candidates, VLA-38, matches the position of a newly detected redback MSP, Ter 5 ar \citep{Padmanabh2024,Corcoran2024}, as well as the X-ray source CX19. This leaves two spider pulsars (Ter 5 aq and Ter 5 at) in Terzan 5 that do not have previously published X-ray counterparts.  

The present study leverages over a decade of \textit{Chandra} observations of Terzan 5 to conduct a detailed photometric, spectral, and variability analysis of more than 100 X-ray sources located within 2'x2' of the cluster center.
% continue here the section numbers etc...

\section{Observations and Data Reductions} \label{sec:obs}

The X-ray data utilized in this study comprised 18 Chandra X-ray Observatory observations of Terzan 5, totaling an exposure time of 736.9 ks (see Table \ref{tab:obs}). For all these observations, the core of Terzan 5 was positioned on the back-illuminated \texttt{ACIS-S3} chip and configured in \texttt{FAINT} mode. We performed data reduction and analysis using \texttt{CIAO}\footnote{Chandra Interactive Analysis of Observations; available at \href{https://cxc.cfa.harvard.edu/ciao/}{https://cxc.cfa.harvard.edu/ciao/}} (version 4.16.0, CALDB 4.11.5), provided by the Chandra X-ray Center.

The data were reprocessed using the \texttt{chandra\_repro} script, which generated new level 2 event files for each observation, incorporating the latest calibration updates and bad pixel files. We performed background filtering in the energy range 0.5–8 keV with a bin time of 100 seconds using the \texttt{dmgti} command. The background was defined to exclude the 1-arcmin region around the cluster.

Observation 3798 exhibited a strong flare towards the end of the light curve, so we removed 8.7 ks of data. The light curves for observations 13252 and 13705 contained some data points abnormally higher than the rest. For observation 13252, this resulted in the removal of only 0.2 ks of data, while for Observation 13705, background filtering did not significantly affect the exposure time. For the remaining 15 observations, no background filtering was required.

To align the observations while correcting for astrometry, we used the reference point as the subarcsecond location of IGR J17480-2446 \citep{Riggio2012}. We matched the observed location of this quiescent source in each observation with its actual position to determine the offsets using \texttt{wcs\_match} and applied them to each event file and aspect solution file using \texttt{wcs\_update}. All event files were then merged using the \texttt{reproject\_obs} script.

\begin{table}
\caption{\textit{Chandra}/\texttt{ACIS-S3} observations of Terzan 5. The exposures with an ``$*$'' were affected by background filtering.}
\label{tab:obs}
    \centering
    \begin{tabular}{ccc}
        \toprule
        Date of Observation & Exposure (ks) & Observation ID\\
        \toprule
         2003 Jul 13 & 30.6$^*$ & 3798 \\
         2009 Jul 15 & 36.3 & 10059 \\
         2011 Feb 07 & 29.7 & 13225 \\
         2011 Apr 29 & 39.3$^*$ & 13252 \\
         2011 Sep 05 & 13.9$^*$ & 13705 \\
         2011 Sep 08 & 34.1 & 14339 \\
         2012 May 13 & 46.5 & 13706 \\
         2012 Sep 17 & 30.5 & 14475 \\
         2012 Oct 28 & 28.6 & 14476 \\
         2013 Feb 05 & 28.6 & 14477 \\
         2013 Feb 22 & 49.2 & 14625 \\
         2013 Feb 23 & 84.2 & 15615 \\
         2013 Jul 16 & 28.6 & 14478 \\
         2014 Jul 15 & 28.6 & 14479 \\
         2014 Jul 17 & 71.6 & 16638 \\
         2014 Jul 20 & 23.0 & 15750 \\
         2016 Jul 13 & 68.9 & 17779 \\
         2016 Jul 15 & 64.7 & 18881 \\
        \hline 
    \end{tabular}
\end{table}

\subsection{Source Detection} \label{sec:sourcedet}

\begin{figure*}
    \centering
    \begin{minipage}{0.49\textwidth}
        \centering
        \includegraphics[width=\linewidth]{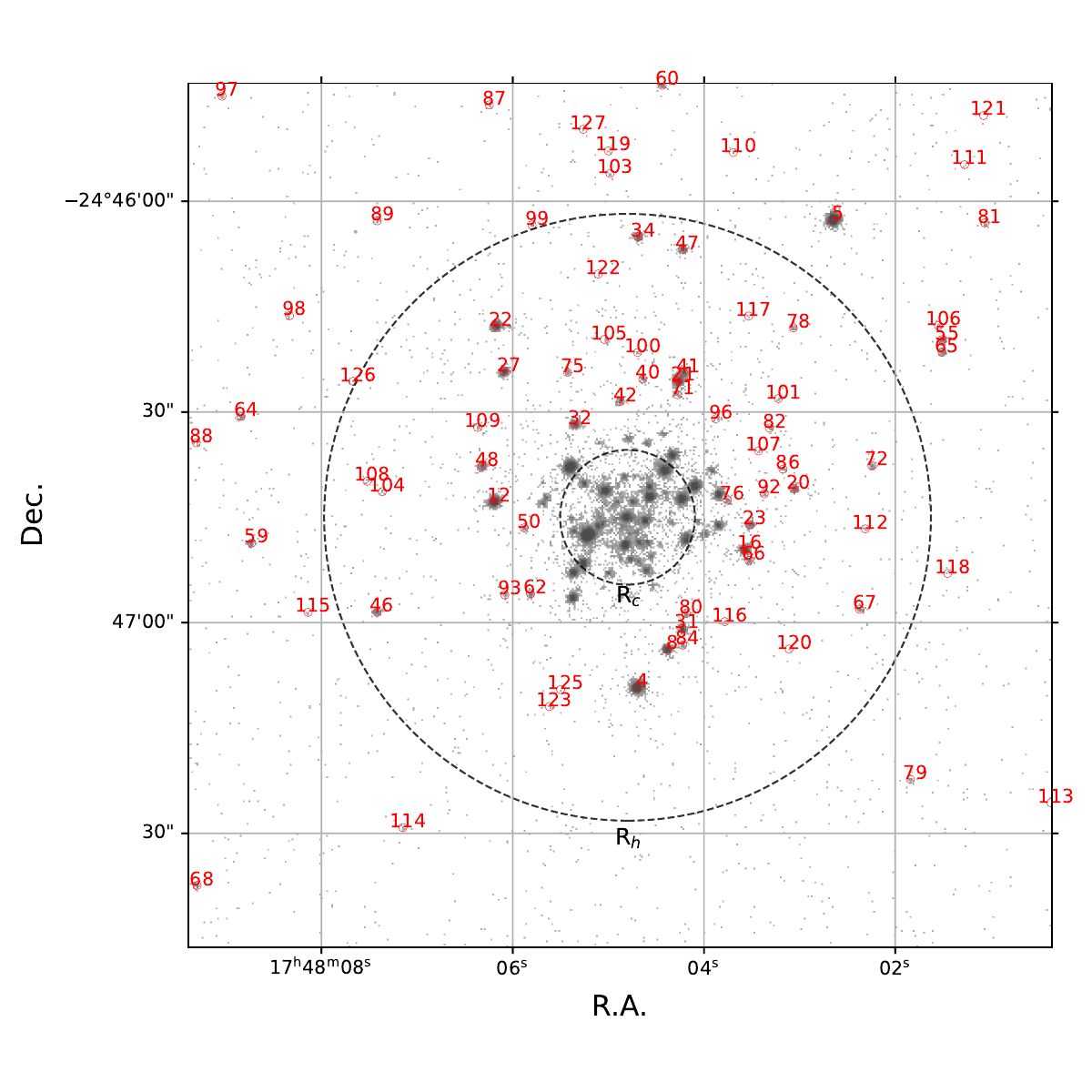} % Second plot
    \end{minipage}
    \begin{minipage}{0.48\textwidth}
        \centering
        \includegraphics[width=\linewidth]{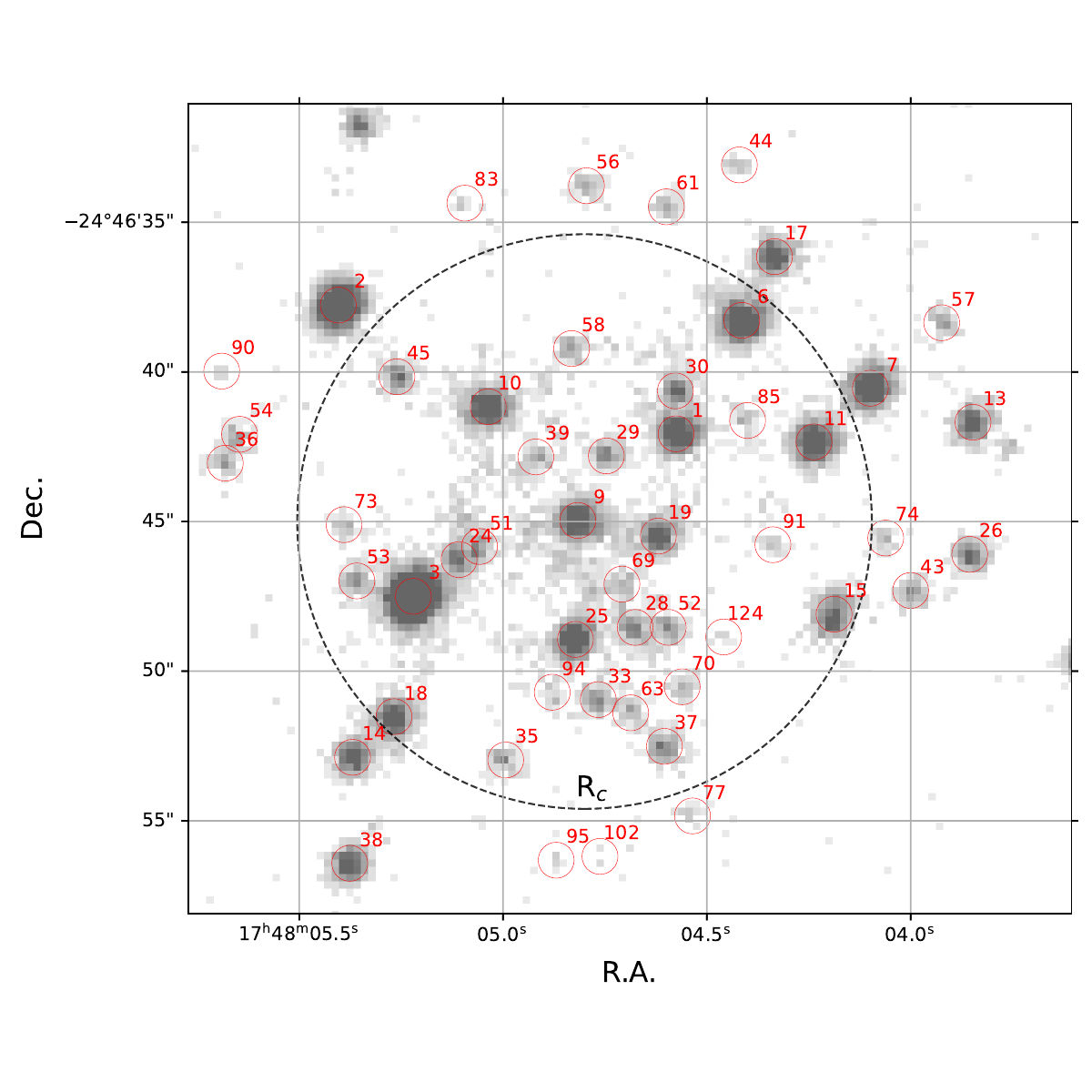} % First plot
    \end{minipage}%
    \caption{Merged Chandra ACIS-S 0.5–8 keV images of Terzan 5. The 126 detected X-ray sources are marked with small (red) circles of 0.6" radius and labelled with their CX IDs \citep{Heinke2006}. The inner black circle represents the core radius, R$_c$ (0.18'), and the larger black circle represents the half-mass radius, R$_h$ (0.72'), of the globular cluster Terzan 5. The left panel shows the full region considered (with sources outside the core labelled), while the right panel shows and labels the detections in and just outside the core. }
    \label{fig:XrayImage}
\end{figure*}

\begin{figure}
    \centering
    \includegraphics[scale=1.1]{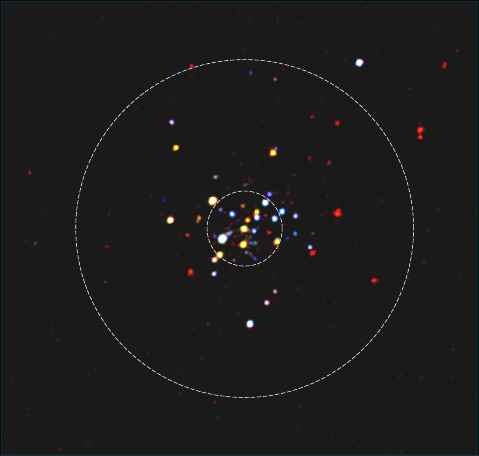}
    \caption{False-color image of Terzan 5. The red, green, and blue colors represent the X-ray counts in the 0.5-1.2, 1.2-2, and 2-8 keV energy bands, respectively. The inner white circle marks the core radius, R$_c$ (0.18'), and the larger black circle denotes the half-mass radius, R$_h$ (0.72'), of the globular cluster Terzan 5.}
    \label{fig:XrayImageRGB}
\end{figure}

We utilized the \texttt{CIAO} package \texttt{wavdetect} for the detection of X-ray sources within 2'x2' around the cluster's center. The wavelet scales were specified as a series from 1 to 8, increasing by a factor of $\sqrt{2}$, and the significance threshold for source detection was set at $10^{-6}$ (false sources per pixel). The data were filtered into three distinct energy ranges: 0.5–2, 2–8, and 0.5–8 keV. For each energy range, the same wavelet scales were employed. The unique X-ray source detections from each energy range were subsequently combined, resulting in a master file containing 126 X-ray sources (see Table \ref{tab:XraySources}). The X-ray sources are shown in Figure \ref{fig:XrayImage}. The RA and DEC of the X-ray source counterparts are corrected to best match the radio positions of EXO 1745-248 \citep{Tetarenko2016}, MSP P, and MSP ad \citep{Prager2017, Bogdanov2021}. 
The systematic errors associated with this alignment procedure in RA and DEC are \(0.002''\) and \(0.1''\), respectively.
Figure \ref{fig:XrayImageRGB} shows a false-color image of Terzan 5, where red, green, and blue represent the X-ray counts in the 0.5–1.2, 1.2–2, and 2–8 keV energy bands, respectively.

We found that 125 of our sources are also detected by \citet{Cheng2019} (CX~125 is absent from their catalog). They also reported an additional 80 sources within a $2\arcmin \times 2\arcmin$ region around the cluster center, which we have not detected. Upon closer inspection, we found some of their sources to be plausible, while the majority are less confident detections. \citet{Bahramian2020} cataloged 212 sources within 1.2 times the half-light radius, including a large number of sources that they characterize as marginal detections. A total of 106 of our sources are also detected by them, while our remaining 20 sources are either not detected by them or fall outside their search radius. They performed source detection using \texttt{PWDetect} \citep{Damiani1997}, which enhances the detection of faint sources in crowded regions. Since the goal of our study is to identify brighter sources suitable for spectral analysis, we adopted a more conservative detection threshold (\texttt{wavdetect sigthresh:} 10$^{-6}$) that excludes fainter sources, and we do not attempt to determine which of the fainter sources are more reliable. However, a \texttt{wavdetect} \texttt{sigthresh} of \(10^{-5}\) was used for some of the fainter sources to determine their counts in individual energy bands.

\subsection{Spectral Extraction} \label{sec:specextract}

We crossmatched our 126 X-ray sources with the Terzan 5 X-ray catalog of \cite{Bahramian2020}. For the successful cross-matches, the corresponding Bahramian source IDs for our X-ray sources are also provided in Table \ref{tab:XraySources}. 
The spectra used in this study were taken directly from \citet{Bahramian2020}, which also provides a detailed description of their spectral extraction procedure using the \textsc{ACIS-Extract} software \citep{Broos2010}. We did not perform spectral extraction in this work for sources not detected by \citet{Bahramian2020}.
We noticed that the spectra of the faint sources in the core from \cite{Bahramian2020} clearly underestimate the background. \citet{Bahramian2020} used \textsc{ACIS-Extract}'s option for model-based background extraction, as advised for crowded regions, but the density of sources (including marginal candidate detections) in the core was so large that constructing appropriate local background regions was not possible. To resolve this issue, we changed the \texttt{AREASCAL} parameter in each of the source background spectra files by a factor of B$_{bkg}$/G$_{bkg}$, where B$_{bkg}$ is the estimated background counts in the source region from \cite{Bahramian2020} and G$_{bkg}$ is the estimated background counts from our source detection analysis (\texttt{wavdetect} $^*src.fits$ output).

\section{X-ray Color-Magnitude Diagram} \label{sec:xraycmd}

\begin{figure*}
    \centering
    \includegraphics[scale=0.8]{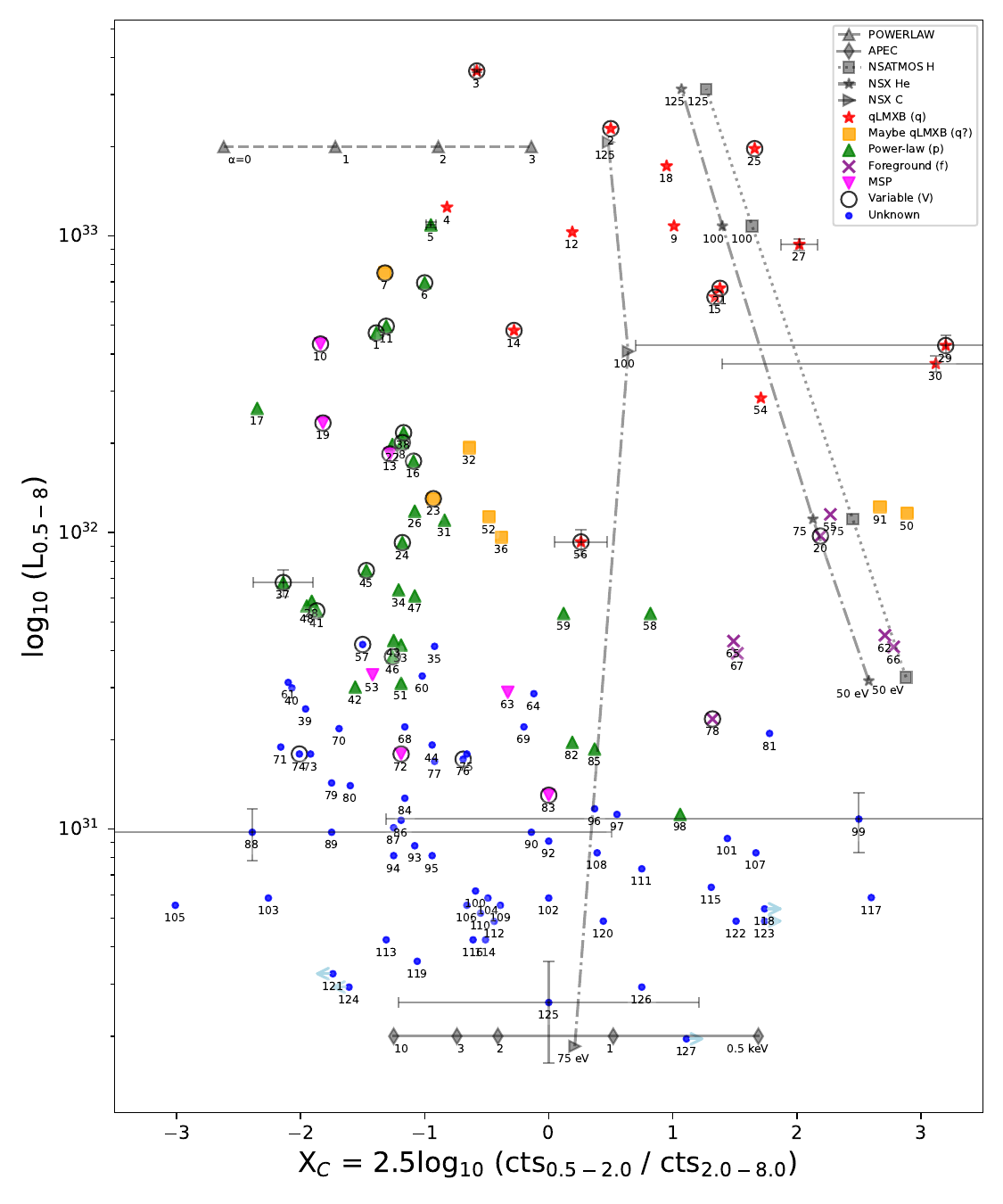}
    \caption{X-ray CMD, plotting X-ray luminosity (0.5–8 keV) against hardness (increasing to the left) for Terzan 5 X-ray sources. All the X-ray sources are annotated with their corresponding CX IDs. qLMXBs (q) are marked with red stars, possible qLMXBs (q?) are marked with orange squares, MSPs are marked with magenta downward triangles, foreground (f) sources are marked with purple crosses, and remaining sources that are best fit with a power-law (p) are marked with green triangles.  All the variable (V) sources are encircled in black. The unknown sources, for which spectral analysis was not specifically conducted, are indicated by blue circles. The predicted locations of several model spectra are indicated, assuming the cluster's $N_H$ column. The spectra are labeled with units of temperature (for H, He and C atmospheres in eV and APEC in keV), and photon index (for the power-law). The vertical locations of the APEC and power-law models are arbitrary. Only a few error bars are plotted, to improve readability. For some of the fainter sources, only an upper limit or lower limit for the colors was estimated, as indicated by light blue arrows.}
    \label{fig:xray_cmd}
\end{figure*}

We created an X-ray CMD (Figure \ref{fig:xray_cmd}) by plotting broadband luminosity, $log_{10}(L_{0.5-8})$ versus the hardness ratio, $X_C = 2.5 log_{10}(cts_{0.5-2.0} / cts_{2.0-8.0})$. To calculate the luminosity, the 50 brightest sources were categorized into qLMXBs and non-qLMXBs (see \S~\ref{sec:spectralanalysis}). For qLMXBs, the fitted NSATMOS plus power-law luminosities (when both components were needed) were added together, while for non-qLMXBs, the power-law luminosity was applied. For the remaining sources, the luminosities were interpolated as power-law luminosities for hard sources ($X_C > 0$) and NSATMOS plus power-law luminosities for soft sources ($X_C < 0$). 

Most sources had count detections in both the 0.5–2.0 keV and 2.0–8.0 keV bands, making the $X_C$ calculation straightforward. However, for some of the fainter sources, where {\tt wavdetect} did not detect the source in one band, the corresponding counts in that band were estimated as the difference between the total 0.5–8.0 keV counts and the counts detected in the other band. For CX118, CX121, CX123, CX124, and CX127, we found zero counts in one of the bands, meaning only an upper limit or lower limit for the colors could be determined. The errors in luminosity were estimated by directly scaling the errors in $cts_{0.5-8.0}$ (these are thus lower limits on the errors, as they do not account for systematic uncertainties in the spectral shape). The errors in colors were derived from the count errors in the different energy bands. The luminosity and $X_C$ values with errors for all the sources are provided in Table \ref{tab:XraySources}.

We note in Fig.~\ref{fig:xray_cmd} the locations of several model spectra, assuming the cluster's $N_H$ column (we use $N_H=2.3\times10^{22}$ cm$^{-2}$; see \S \ref{sec:spectralanalysis}). The spectra are labeled with units of temperature (for H, He and C atmospheres in eV and APEC in keV) and photon index (for the power-law). The vertical locations of the APEC and power-law models are arbitrary. Through spectral analysis (see \S~\ref{sec:spectralanalysis}), we classified many sources as qLMXBs, possible qLMXBs, power-law sources, and foreground sources, as annotated in Figure \ref{fig:xray_cmd}. We also identify sources that are variable (\S~\ref{sec:variabilityanalysis}) with black circles in Fig.~\ref{fig:xray_cmd}. 

\begin{figure}
    \centering
    \includegraphics[scale=0.65]{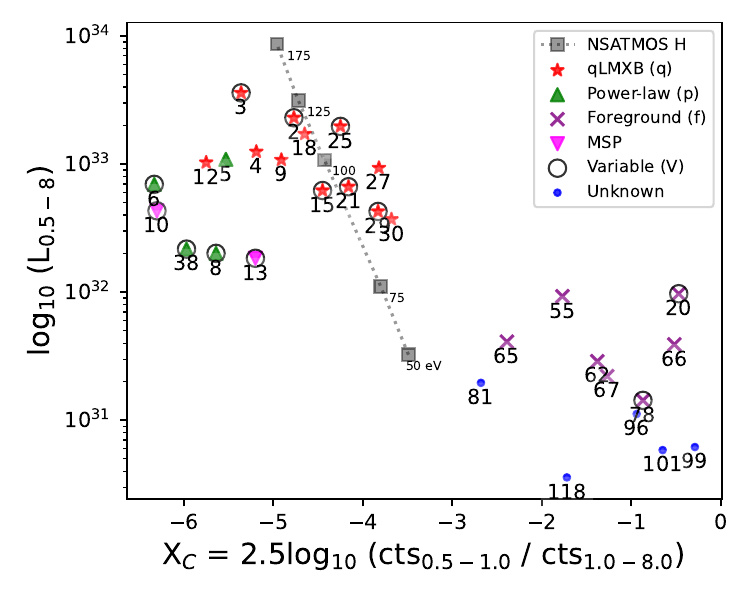}
    \caption{X-ray color-magnitude diagram, using 0.5-1.0 vs. 1.0-8.0 keV, only plotting sources detected in 0.5-1.0 keV. This clearly separates the verified foreground stars from qLMXBs and other sources in Terzan 5.}
    \label{fig:xraycmd_0.5-1}
\end{figure}

We also created a second X-ray CMD using bands selected to separate foreground sources from qLMXBs; 0.5-1.0 keV and 1.0-8.0 keV.  Fig.~\ref{fig:xraycmd_0.5-1} shows a clear separation between the sources in Terzan 5 (including qLMXBs) and the foreground sources, with lower $N_H$, which thus have much higher ratios of flux in 0.5-1.0 keV to 1.0-8.0 keV. In \S~\ref{sec:spectralanalysis}, we use spectral fitting to identify the likely foreground sources which have $>$50 counts. We also notice in Fig.~\ref{fig:xraycmd_0.5-1} several fainter sources that are candidate foreground sources, but with too few counts for spectral fitting.

\section{Variability Analysis} \label{sec:variabilityanalysis}

We performed variability analysis on all 126 X-ray sources, searching for both long-term variability and variability within individual observations. We used the \texttt{dmstat} command to estimate the counts for each source in each observation. To include error bars in the light curve, we applied 1-$\sigma$ Poisson statistics using Equation 7 from \cite{Gehrels1986}. The $\chi^2$ statistic was calculated using Equation \ref{eq:chisquare}:

\begin{equation} \label{eq:chisquare} \chi^2 = \sum \limits_{\text{sum over obs.}} \frac{(\text{counts} - \text{avg count rate} \times \text{exposure})^2 }{(\sqrt{\text{counts}+\frac{3}{4}}+1)^2} \end{equation}
We classified a source as variable if $\chi^2 > 33$ (99\% confidence interval, for 18 observations).

We also checked for short-term variability, by checking whether the \texttt{PROB\_KS} parameter produced by \textsc{ACIS-Extract} \citep{Bahramian2020, Broos2010} is $<$ 0.001 in \texttt{source.stats} for any observation of an X-ray source. The \texttt{PROB\_KS} parameter provides the probability that an observation of a randomly-generated nonvariable lightcurve 
would have a KS statistic higher than the observed value. We chose a threshold of 0.001 to give 1 expected false detection when accounting for the total number of trials. Using this criterion, we identified 30 variable sources (see notes in Table \ref{tab:XraySources}). 

We verified variability within individual observations for sources that had \texttt{PROB\_KS} $< 0.001$ but $\chi^2 < 33$, using \texttt{CIAO}'s Gregory-Loredo variability test algorithm (\texttt{glvary}), and the results agreed with the \texttt{PROB\_KS} analysis. CX6, CX11, CX21, CX29, and CX57 exhibited short-term variability. CX74 displayed two counts 70 seconds apart in observation 3798, while CX78 exhibited six counts within 1400 seconds in observation 14625.

\section{Spectral Analysis} \label{sec:spectralanalysis}

We utilized \texttt{CIAO}'s modeling and fitting application, \texttt{SHERPA}\footnote{\href{https://cxc.cfa.harvard.edu/sherpa/}{https://cxc.cfa.harvard.edu/sherpa/}} for all spectral fits. We grouped the spectral data for the sources using different bin sizes: 100 counts per bin for total counts \(> 1000\), 50 counts per bin for \(> 500\), 20 counts per bin for \(> 200\), 10 counts per bin for \(> 100\), and 1 count per bin for \(< 100\). This was done to ensure optimal sampling of the spectra. Sources with fewer than 100 counts were fitted using \textit{wstat}\footnote{\url{https://heasarc.gsfc.nasa.gov/xanadu/xspec/manual/XSappendixStatistics.html}} statistics. \textit{wstat} is a Poisson maximum-likelihood method appropriate for spectral fitting when both source and background data have Poisson noise.  
It accounts for background without subtracting it directly and works well for low-count statistics.

In all the spectral fits, we set the elemental abundances to \texttt{wilm} \citep{Wilms2000}. To determine the hydrogen column density ($N_H$), we performed spectral fits by keeping $N_H$ free for the first five brightest sources and found that all sources agreed on an approximate $N_H$ of $2.3\times10^{22}$ cm$^{-2}$, which was then used in the spectral fits.

We considered four different sets of models to classify the X-ray sources into different categories.

For a blackbody-like neutron star H-atmosphere model, we used the XSpec photoelectric absorption model (\texttt{xsphabs}) multiplied by the sum of the XSpec power-law with pegged normalization (\texttt{xspegpwrlw}) and the low-magnetic-field NS hydrogen atmosphere model (\texttt{xsnsatmos}, \cite{HeinkeNS2006}), with the entire combination further multiplied by the XSpec super-exponential cutoff absorption model to account for dust scattering dependent on the absorption column (\texttt{xsspexpcut}, \cite{Predehl2003}). Although dust scattering is more significant in cases where \(N_{\mathrm{H}} \ge 10^{23} \, \text{cm}^{-2}\), 
for $N_H$ values of $\sim10^{22}$ cm$^{-2}$ we find that its effects are similar to a few percent increase in $N_H$. For a blackbody-like neutron star He-atmosphere model, we used the same combination as in the H-atmosphere model, except that \texttt{xsnsatmos} was replaced with the non-magnetic He neutron star atmosphere model (\texttt{xsnsx}, \texttt{specfile=2}, \cite{Ho2009}). The \texttt{alpha} parameter of \texttt{xsspexpcut} was fixed to -2, while the cutoff energy (\texttt{Ecut}) was set to $\sqrt{0.49\times N_H [10^{22}\,cm^{-2}]}$ keV \citep{Predehl2003}. The NS gravitational mass (M$_{ns}$) was fixed at 1.4 M$_\odot$, while the NS radius (R$_{ns}$) was fixed at 12 km. The distance to the NS (dist) was set to 6.6 kpc \citep{Baumgardt2021}. The energy range for the \texttt{xspegpwrlw} normalization was chosen to be 0.5–10 keV.

For the power-law spectrum model, the \texttt{xsnsatmos} component was removed while keeping the rest of the components unchanged. Finally, for the hot plasma emission model, we used the APEC emission spectrum (\texttt{xsapec}) multiplied by \texttt{xsphabs} and \texttt{xsspexpcut}. The \texttt{Abundanc} parameter of the \texttt{xsapec} model was fixed to 0.63 ($\approx$ 10$^{-0.2}$, [Fe/H] = -0.2, \cite{Harris1996}). An APEC fit with \(N_{\mathrm{H}}\) close to zero suggests the possibility of foreground sources.

Based on the spectral fitting, we classified the sources into qLMXBs (q), maybe qLMXBs (q?), power-law sources (p), and foreground sources (f). The F-test was performed between $\chi^2$ of the fit for NSATMOS+power-law and the power-law fit. If the F-test probability was 0-5\% (i.e. favoring the NSATMOS+power-law fit), or \texttt{PhoIndex} $>$ 3 for the power-law fit, then the source was classified as a confident qLMXB (q) candidate. However, an F-test probability of 5-10\% or \texttt{PhoIndex} $\approx$ 2-3 suggests a weak NS signature in the spectrum, leading to its classification as maybe qLMXB (q?) candidate. (The probability of the spectrum appearing by chance to require a NS signature is small, but large enough that 1-2 of our candidate qLMXBs may not be real qLMXBs.)  We also attempted to see if the NSX+power-law (He-atmosphere) model provided a better fit than NSATMOS+power-law for qLMXBs, but we could not distinguish between the quality of these fits. Table \ref{tab:qLMXB} presents the NSATMOS+power-law fitting results for the 22 qLMXB/maybe qLMXB (q/q?) sources found in Terzan 5. There are a few q/q? sources for which we had to constrain the \texttt{PhoIndex} in NSATMOS+power-law fitting between 1 and 2.5, to prevent implausible \texttt{PhoIndex} values (these have fixed values of photon index in Table~\ref{tab:qLMXB}). These sources do not formally require (using an F-test) a power-law component in their spectral fits, but we give details from the fit allowing a power-law fit anyway, as qLMXBs may have a power-law spectral component below our sensitivity.
We calculated the fraction of unabsorbed X-ray luminosity coming from the NSATMOS component ($\frac{L_{X,\text{NSATMOS}}}{L_{X,\text{Total}}}$) using \texttt{SHERPA}'s \texttt{calc\_energy\_flux} function on the NSATMOS model while excluding the power-law component.

\begin{table}
\caption{Spectral results of the NSATMOS+power-law fit for the qLMXB/ maybe qLMXB (q/q?) sources. The derived NS surface temperature is given as \texttt{log10(T$_{eff}$)}. The derived photon index of the power-law component is denoted as \texttt{PhoIndex}. The fraction of total X-ray luminosity coming from the NSATMOS component is given as $\frac{L_{X,\text{NSATMOS}}}{L_{X,\text{Total}}}$. Note: Sources with fewer than 100 counts were fitted using \textit{wstat} statistics.}
\label{tab:qLMXB}
    \centering
    \begin{tabular}{ccccc}
        \toprule
        CX & \texttt{log10(T$_{eff}$)} & \texttt{PhoIndex} & $\frac{L_{X,\text{NSATMOS}}}{L_{X,\text{Total}}}$ & \texttt{$\chi_R^2$/dof} \\
        \toprule
2	&	6.160	$_{	-0.003	}^{+	0.002	}$ &	1.1	$_{	-0.2	}^{+	0.2	}$ &	0.82	$_{	-0.03	}^{+	0.02	}$ &	1.01/39	\\
3	&	6.146	$_{	-0.007	}^{+	0.007	}$ &	1.4	$_{	-0.1	}^{+	0.1	}$ &	0.45	$_{	-0.03	}^{+	0.04	}$ &	0.41/86	\\
4	&	5.99	$_{	-0.03	}^{+	0.02	}$ &	1.84	$_{	-0.09	}^{+	0.09	}$ &	0.24	$_{	-0.07	}^{+	0.06	}$ &	1.23/34	\\
7	&	5.95	$_{	-0.05	}^{+	0.03	}$ &	1.2	$_{	-0.1	}^{+	0.1	}$ &	0.19	$_{	-0.05	}^{+	0.19	}$ &	0.86/19	\\
9	&	6.08	$_{	-0.008	}^{+	0.01	}$ &	2	$_{	-2	}^{+	2	}$ &	0.85	$_{	-0.08	}^{+	0.09	}$ &	0.41/11	\\
12	&	6.072	$_{	-0.007	}^{+	0.006	}$ &	1.6	$_{	-0.2	}^{+	0.2	}$ &	0.75	$_{	-0.05	}^{+	0.05	}$ &	0.54/15	\\
14	&	5.99	$_{	-0.02	}^{+	0.01	}$ &	1.8	$_{	-0.2	}^{+	0.2	}$ &	0.6	$_{	-0.1	}^{+	0.1	}$ &	1.34/16	\\
15	&	6.04	$_{	-0.008	}^{+	0.006	}$ &	1.6	$_{	-0.5	}^{+	0.5	}$ &	0.88	$_{	-0.07	}^{+	0.06	}$ &	0.53/15	\\
18	&	6.063	$_{	-0.005	}^{+	0.006	}$ &	2.5	$_{	-0.9	}^{		}$ &	0.91	$_{	-0.05	}^{+	0.07	}$ &	0.51/7	\\
21	&	6.05	$_{	-0.007	}^{+	0.005	}$ &	1	$_{	-1	}^{+	1	}$ &	0.93	$_{	-0.07	}^{+	0.05	}$ &	0.85/12	\\
23	&	5.84	$_{	-0.08	}^{+	0.03	}$ &	1.4	$_{	-0.4	}^{+	0.4	}$ &	0.6	$_{	-0.4	}^{+	0.2	}$ &	0.59/18	\\
25	&	6.102	$_{	-0.004	}^{+	0.003	}$ &	2	$_{		}^{		}$ &	1.00	$_{	-0.04	}^{}$ &	0.55/12	\\
27	&	6.03	$_{	-0.01	}^{+	0.006	}$ &	1	$_{		}^{		}$ &	0.96	$_{	-0.11	}^{+	0.04	}$ &	0.75/6	\\
29	&	5.972	$_{	-0.007	}^{+	0.006	}$ &	2	$_{		}^{		}$ &	1.00	$_{	-0.08	}^{}$ &	0.59/23	\\
30	&	6.003	$_{	-0.008	}^{+	0.006	}$ &	1	$_{		}^{		}$ &	1.00	$_{	-0.09	}^{}$ &	0.74/12	\\
32	&	5.89	$_{	-0.06	}^{+	0.03	}$ &	1.9	$_{	-0.4	}^{+	0.3	}$ &	0.54	$_{	-0.3	}^{+	0.2	}$ &	0.67/16	\\
36	&	5.8	$_{	-0.1	}^{+	0.05	}$ &	2	$_{	-1	}^{+	1	}$ &	0.57	$_{	-0.48	}^{+	0.06	}$ &	0.50/8	\\
50	&	5.904	$_{	-0.009	}^{+	0.007	}$ &	1	$_{		}^{		}$ &	1.0	$_{	-0.1	}^{}$ &	0.93/56	\\
52	&	5.87	$_{	-0.07	}^{+	0.03	}$ &	1.3	$_{	-0.8	}^{+	0.8	}$ &	0.7	$_{	-0.4	}^{+	0.3	}$ &	0.39/15	\\
54	&	5.93	$_{	-0.01	}^{+	0.01	}$ &	2.5	$_{		}^{		}$ &	0.97	$_{	-0.14	}^{+	0.03	}$ &	0.09/8	\\
56	&	5.88	$_{	-0.05	}^{+	0.03	}$ &	2	$_{	-2	}^{+	2	}$ &	0.8	$_{	-0.4	}^{+	0.2	}$ &	0.41/7	\\
91	&	5.90	$_{	-0.01	}^{+	0.01	}$ &	1	$_{	-2	}^{		}$ &	0.96	$_{	-0.15	}^{+	0.04	}$ &	0.90/56	\\
        \hline 
    \end{tabular}
\end{table}

Eq.~\ref{eq:redchisq} defines the reduced chi-square (\(\chi^2_R\)), where \(O_i\) and \(E_i\) are the observed and expected (model) values respectively, \(\sigma_i\) is the uncertainty in the observed value, and \(R = N - p\) is the number of degrees of freedom, with \(N\) being the number of data points and \(p\) the number of fitted parameters. 

\begin{equation} \label{eq:redchisq}
\chi^2_R = \frac{1}{R} \sum_{i=1}^{N} \left( \frac{O_i - E_i}{\sigma_i} \right)^2
\end{equation}

Sources where the $\chi_R^2$ for the power-law fit was better or comparable to the NSATMOS+power-law fit were classified as power-law sources. If none of the models provided good fits, we tried an APEC fit with $N_H$ free. For six sources, we then obtained good fits, with the  derived $N_H$ inconsistent with the cluster value, typically close to zero. This suggests that these sources are not cluster members; hence, we classified them as foreground sources.

\section{Radial Distribution Analysis} \label{sec:radial_analysis}

\newcommand{\chandra}{\emph{Chandra}}

Following the approach of our previous studies of the radial distributions of subclasses of \chandra\ sources in clusters \citep[e.g.][]{Heinke2006,Lugger2007,Lugger2017, Lugger2023,Cohn2010,Cohn2021,RiveraSandoval2018CV}, which has also been applied to Ter 5 by \citet{Prager2017} and \citet{Cheng2019}, we compared the spatial distribution of the qLMXB candidates and the 50-count CX (\textit{Chandra} X-ray) source sample with that of the main sequence turnoff (MSTO) stars. We adopted the King model fit of \citet{Lanzoni2010} as the best description of the surface density profile of the MSTO sample. The core radius of this fit is $r_c = 9\farcs0$ and the concentration parameter is $c=1.49$. Our analysis uses a ``generalized King model'' approximation\footnote{This form, which we used in \citet{lugger1995}, was introduced by \citet{Elson1987} to fit the profiles of young star clusters and has been called the EFF model. It is a generalization of Eqn.~13 of \citet{King1962}.} for the surface density profile,
\begin{equation}
\label{eqn:Generalized_King_Model} 
S(r) = S_0 \left[1 + \left({\frac{r}{r_0}}\right)^2 \right]^{\alpha/2},
\end{equation}
with the core radius $r_c$ related to the scale radius $r_0$ by
\begin{equation}
r_c = \left(2^{-2/\alpha} -1 \right)^{1/2} r_0.
\end{equation}
This model best fits the inner region of a $c=1.49$ King model with the slope parameter set to $\alpha = -2.2$. As discussed in \citet[][see Eqn.~13 therein]{Cohn2021}, thermal equilibrium implies that, for each mass group of typical mass $m$, 
\begin{equation}
\label{eqn:alpha-q_relation}
\alpha = q (\alpha_\mathrm{to}-1) + 1,
\end{equation}
with $q = m/m_\mathrm{to}$.

\begin{figure}
    \centering
    \includegraphics[scale=0.7]{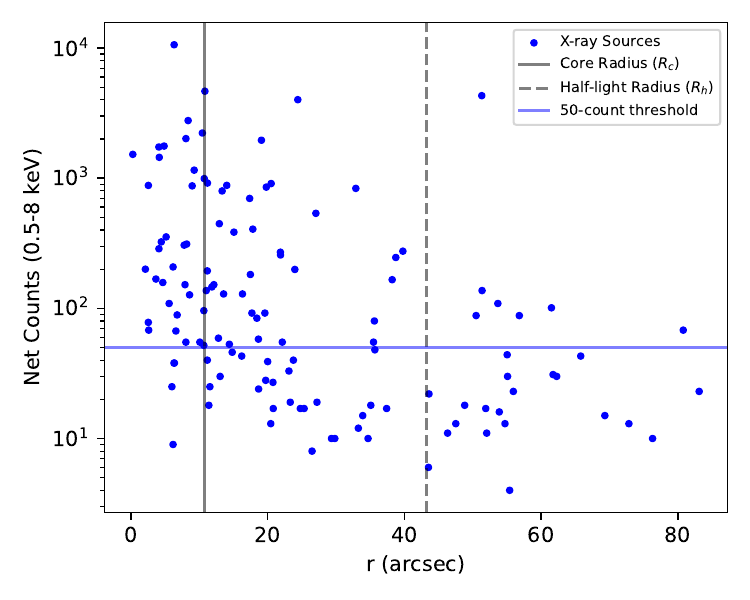}
    \caption{Plot of X-ray source counts versus radial offset from the cluster center. The core radius ($R_c$), half-light radius ($R_h$), and the 50-count threshold are annotated.}
    \label{fig:src_dist}
\end{figure}

\begin{figure}
    \centering
    \includegraphics[width=0.46\textwidth]{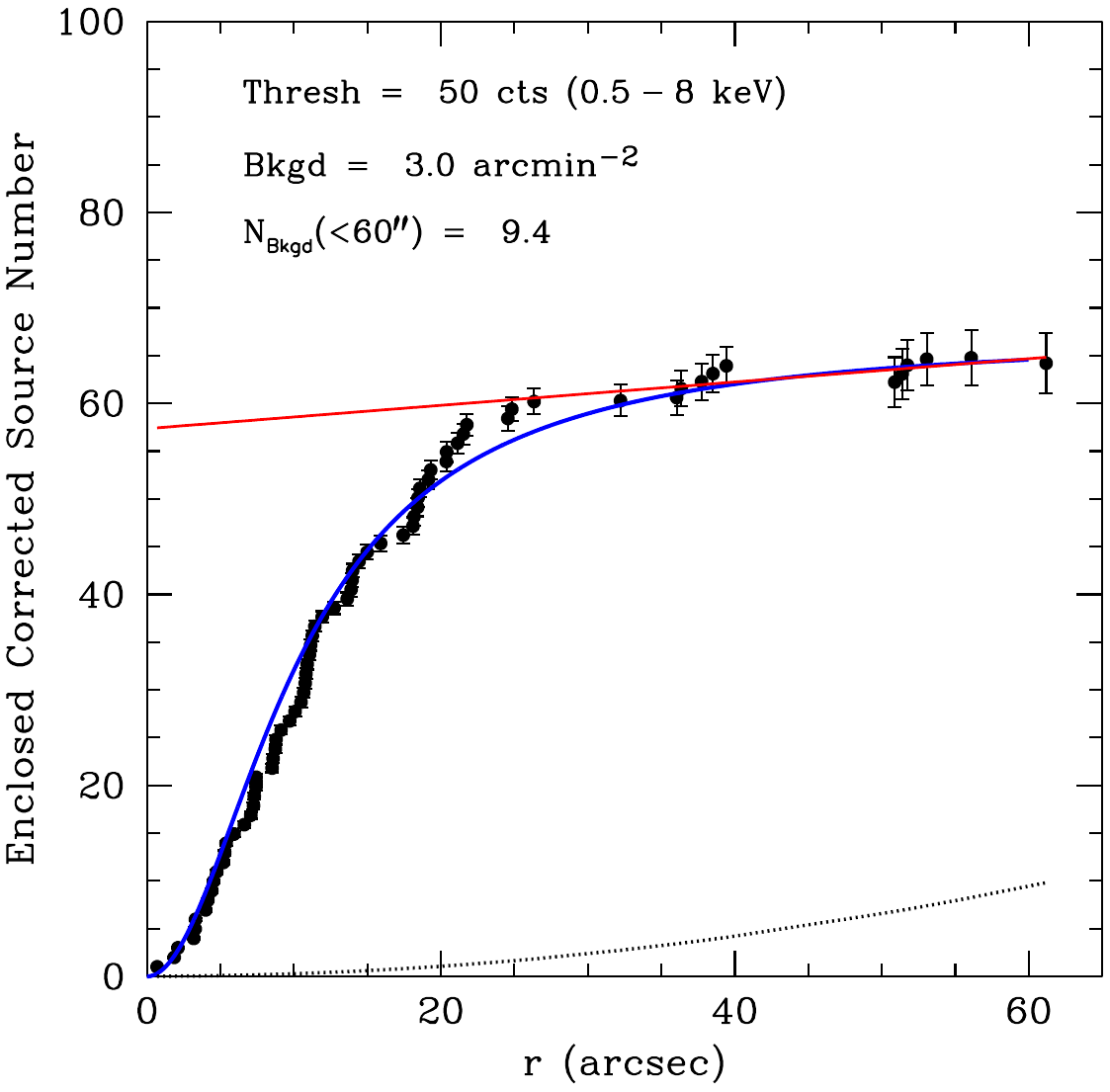}
    \caption{Enclosed source number vs.\ radial offset from cluster center, corrected for foreground/background sources. The error bars represent the statistical uncertainty of the enclosed net source count at each radius. The red line is a linear regression to the corrected counts for $r > 30''$. The blue curve is the maximum-likelihood fit to the cumulative source counts shown in Fig.~\ref{fig:radial_profile}. The dotted line represents the enclosed foreground source number, assuming a uniform distribution. The predicted number of foreground sources within the half-mass radius is 4.1 (\S \ref{sec:foreground&background}).}
    \label{fig:bkgd_correction}
\end{figure}

A plot of net counts for each source vs.\ radial offset from the cluster center indicates some incompleteness for sources with less than 50 counts (Figure \ref{fig:src_dist}).
This diagram shows that the majority of sources between the core and half-light radius have less than 50 counts, while there are few such sources inside the core radius--this is likely caused by crowding. We assume that the true number of sources continues to increase at lower luminosities, allowing a rough identification of an incompleteness limit at the number of net counts below which the number of sources begins to fall.
We therefore restricted our analysis to sources above 50 counts. We performed maximum-likelihood fits of Eqn.~\ref{eqn:Generalized_King_Model} to both the 50-count source sample and the 50-count qLMXB sample. In order to fit the 50-count source sample, it is necessary to make a correction for foreground and background sources. This was done by determining the surface density of 50-count sources beyond the half-mass radius, resulting in a non-cluster correction of 3.0 sources arcmin$^{-2}$. The effect of this correction is illustrated by Fig.~\ref{fig:bkgd_correction}. Note that the net source counts profile appears to become asymptotically flat for large radial offset, indicating that beyond approximately $1'$ there are very few cluster sources. 

\begin{figure}
    \centering
    \includegraphics[width=0.46\textwidth]{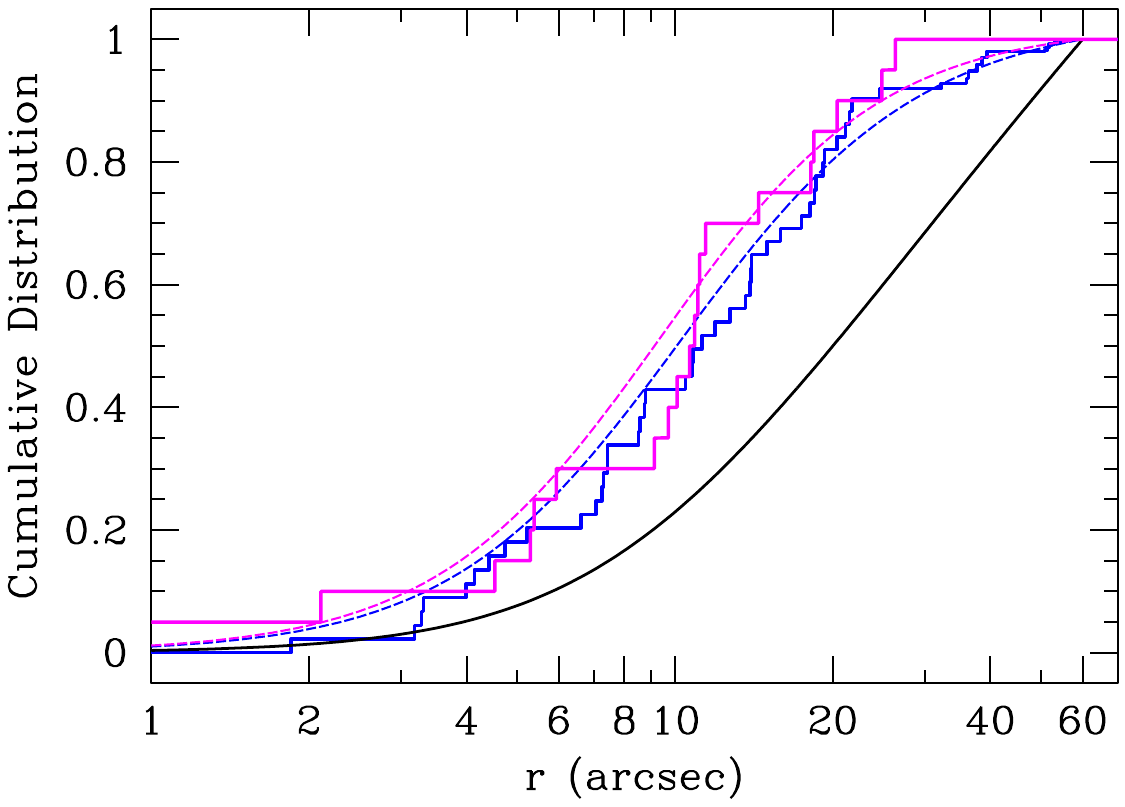}
    \caption{Stair-step curves: cumulative source number vs.\ radial offset from cluster center. Smooth dashed curves: maximum-likelihood fits. Blue curves: $>$50-count source sample, corrected for foreground/background sources. Magenta curves: $>$50-count qLMXB sample. The black curve is the generalized King model that best fits the cluster surface density profile. }
    \label{fig:radial_profile}
\end{figure}

\newcommand{\nd}{\nodata}
\begin{deluxetable*}{lrccccc}
\tablewidth{0pt}
\tablecaption{Generalized King Model Fit Results}
\label{tab:King_Model_Fits}
\tablehead{
\colhead{Sample} &
\colhead{$N$\tablenotemark{a}} &
\colhead{$q$} &
\colhead{$r_c$ (arcsec)} &
\colhead{$\alpha$} &
\colhead{$m\,(\msun)$} &
\colhead{$\sigma$\tablenotemark{b}}
}
\startdata
MSTO	&	\nd	&	1.0	&	$9.0$	&	$-2.2$	&	$0.92 \pm 0.05$	&	\nd	\\
CX	&	64	&	$1.49 \pm 0.11$	&	$6.4 \pm 0.4$	&	$-3.78 \pm 0.35$	&	$1.37 \pm 0.11$	&	3.7	\\
qLMXB	&	20	&	$1.59 \pm 0.14$	&	$6.1 \pm 0.4$	&	$-4.09 \pm 0.45$	&	$1.46 \pm 0.14$	&	3.7	\\
\enddata
\tablenotetext{a}{Size of 50-count sample within $1'$ of cluster center. CX sample is corrected for foreground/background.}
\tablenotetext{b}{Significance of mass difference from MSTO mass.}
\end{deluxetable*}

\newcommand{\ergs}{\mbox{${\rm erg}~{\rm s}^{-1}$}}
With this determination of the foreground/back\-ground level we next performed a maximum-likelihood fit of Eqn.~\ref{eqn:Generalized_King_Model} to the 50-count source sample that lies within $1'$ of the cluster center. This involves determining the single parameter $q_X$ that maximizes the likelihood. The result of this fit is illustrated in Fig.~\ref{fig:radial_profile} and tabulated in Table~\ref{tab:King_Model_Fits}. The best-fit value of the mass ratio is $q_X = 1.49 \pm 0.11$. If we take a value of $0.92\pm0.05 \,\msun$ for the MSTO mass \citep[]{Lanzoni2010,Ferraro2016,Cheng2019}, the implied value of $M_X$ is $1.37 \pm 0.11\,\msun$. For comparison, our previous study of the \chandra\ source distribution in Ter 5 found $q_X = 1.43 \pm 0.11$ for a distribution of 40 10-count sources \citep{Heinke2006}, which is in excellent agreement with the present result. Our present result is also statistically consistent with the ``bright source'' mass in Ter 5 of $M_X = 1.48 \pm 0.11\,\msun$ obtained by \citet{Cheng2019} using a similar maximum likelihood fitting method to that used in this study.\footnote{\citet{Cheng2019} adopted a threshold of $L_{(X,0.5-8)} \sim 9.5 \times 10^{30}\,\ergs$ to define their bright source sample, which approximately corresponds to a level of 30 counts in this study.}

\newcommand{\subq}{{\rm q}}
We next fit Eqn.~\ref{eqn:Generalized_King_Model} to the radial distribution of 20 50-count qLMXBs within $1'$ of the cluster center. The distribution of these objects is very similar to that of the overall sample of 64 50-count \chandra\ sources as seen in Fig.~\ref{fig:radial_profile}. The derived qLMXB mass parameters for this fit are $q_\subq = 1.59 \pm 0.14$ and $m_\subq = 1.46 \pm 0.14\,\msun$. Thus, the characteristic object mass of the qLMXB sample is consistent with that of the overall 50-count CX sample. In our previous study of a sample of 11 likely qLMXBs plus a transient LMXB in Ter 5 we obtained a similar qLMXB to MSTO mass ratio of $q_\subq = 1.64 \pm 0.25$ \citep{Heinke2006}. 

As an additional check, we considered a sample of 50-count \chandra\ sources which excludes the qLMXBs. After correcting for the expected foreground/background the size of this sample is 44. We obtained a value of $q_X = 1.46 \pm 0.14$, which is very similar to the result for the complete 50-count CX sample.

\section{Discussion}

Terzan 5 has the highest interaction rate among all the Galactic globular clusters. By combining spectral, variability, and photometric analysis, we identified many interesting sources within 2'x2' of the cluster center.

\subsection{qLMXB/maybe qLMXB (q/q?) sources}

We identified 22 qLMXB and 'maybe qLMXB' (q/q?) sources in Terzan 5, which is the largest number of qLMXBs identified in any Galactic globular cluster. The X-ray CMD, figure \ref{fig:xray_cmd}, shows a clear dichotomy among X-ray sources above $L_X=10^{32}$ erg/s, with the majority of sources having X-ray colors consistent with a power-law of photon index $\sim$1-2, and another group of much softer sources that are generally consistent with H or He atmosphere NS models. Some of these softer group are actually foreground stars, but these can be separated by their soft (0.5-1 keV vs. 1-8 keV) X-ray color (Fig.~\ref{fig:xraycmd_0.5-1}) and by spectral fitting. Some of the harder group are qLMXBs where the power-law component is stronger \citep[e.g. EXO 1745-248,][]{Wijnands2005}. 

\begin{figure*}
    \centering
    \begin{minipage}{0.45\textwidth}
        \centering
        \includegraphics[width=\linewidth]{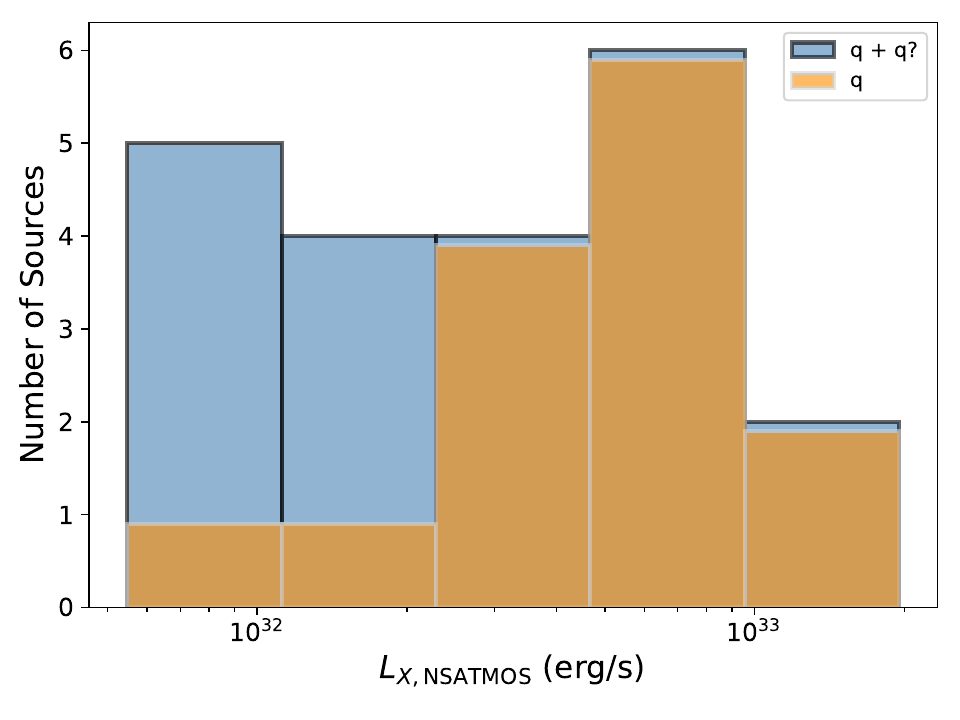} % First plot
    \end{minipage}%
    \begin{minipage}{0.45\textwidth}
        \centering
        \includegraphics[width=\linewidth]{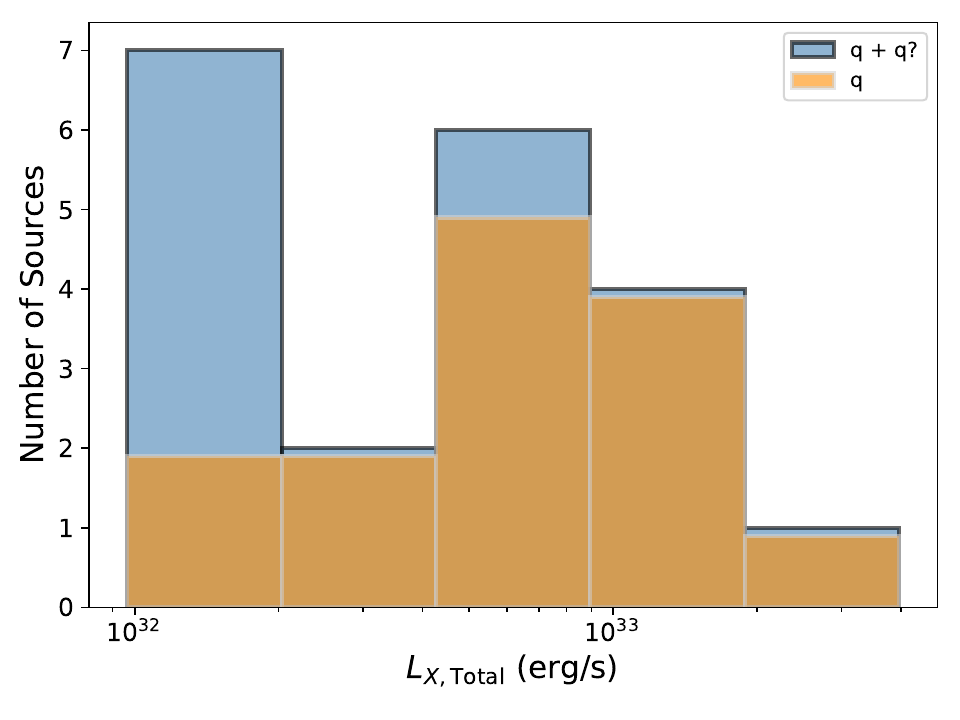} % Second plot
    \end{minipage}
    \caption{Histogram of the unabsorbed X-ray luminosities for confirmed qLMXB (q) sources as well as all q/q? sources. The left panel shows the distribution of unabsorbed X-ray luminosity from the NSATMOS component ($L_{X,\mathrm{NSATMOS}}$), while the right panel shows the distribution of the total unabsorbed X-ray luminosity ($L_{X,\mathrm{Total}}$).}
    \label{fig:qlmxb_lum_hist}
\end{figure*}

Our faintest identified q/q? sources reach an X-ray luminosity of approximately \( 10^{32} \, \text{erg/s} \). Figure \ref{fig:xray_cmd} shows that we would detect a large population of soft qLMXBs down to $L_X\sim3\times10^{31}$ erg/s (foreground sources CX62 and CX66 with similar X-ray colors are still detected with 80-90 counts), but that these do not appear to exist in Terzan 5. Figure \ref{fig:qlmxb_lum_hist} shows the distribution of $L_{X,\text{NSATMOS}}$ and $L_{X,\text{Total}}$ for confident qLMXB (q) sources as well as all q/q? sources.

\begin{figure}
    \centering
    \includegraphics[scale=0.5]{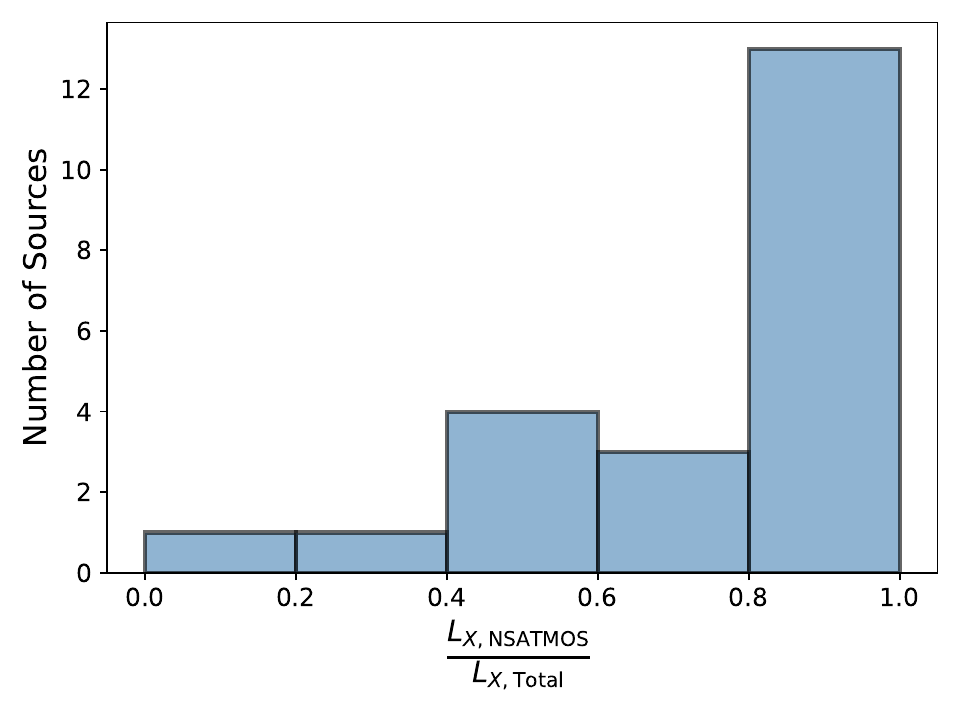}
    \caption{Histogram of the fraction of unabsorbed X-ray luminosity coming from the NSATMOS component ($\frac{L_{X,\text{NSATMOS}}}{L_{X,\text{Total}}}$) for the q/q? sources.}
    \label{fig:ns_frac_hist}
\end{figure}

It is possible that fainter qLMXBs exist in Terzan 5, but present harder spectra; this would fit a pattern of increased relative power-law component in qLMXBs at lower $L_X$ observed by \citet{Jonker04}. Figure \ref{fig:ns_frac_hist} shows the distribution of the fraction of unabsorbed X-ray luminosity coming from the NSATMOS component ($\frac{L_{X,\text{NSATMOS}}}{L_{X,\text{Total}}}$) among the q/q? sources. More than 50\% of these sources have the NSATMOS component contributing more than 80\% to the total luminosity, showing no significant power-law signature. However, three sources have a power-law component that contributes more than the NSATMOS towards the total luminosity. We can only identify such qLMXBs with dominant power-law components among our brighter sources, as we require significant evidence for the NSATMOS component to identify candidate qLMXBs.

\begin{figure}
    \centering
    \includegraphics[scale=0.7]{ 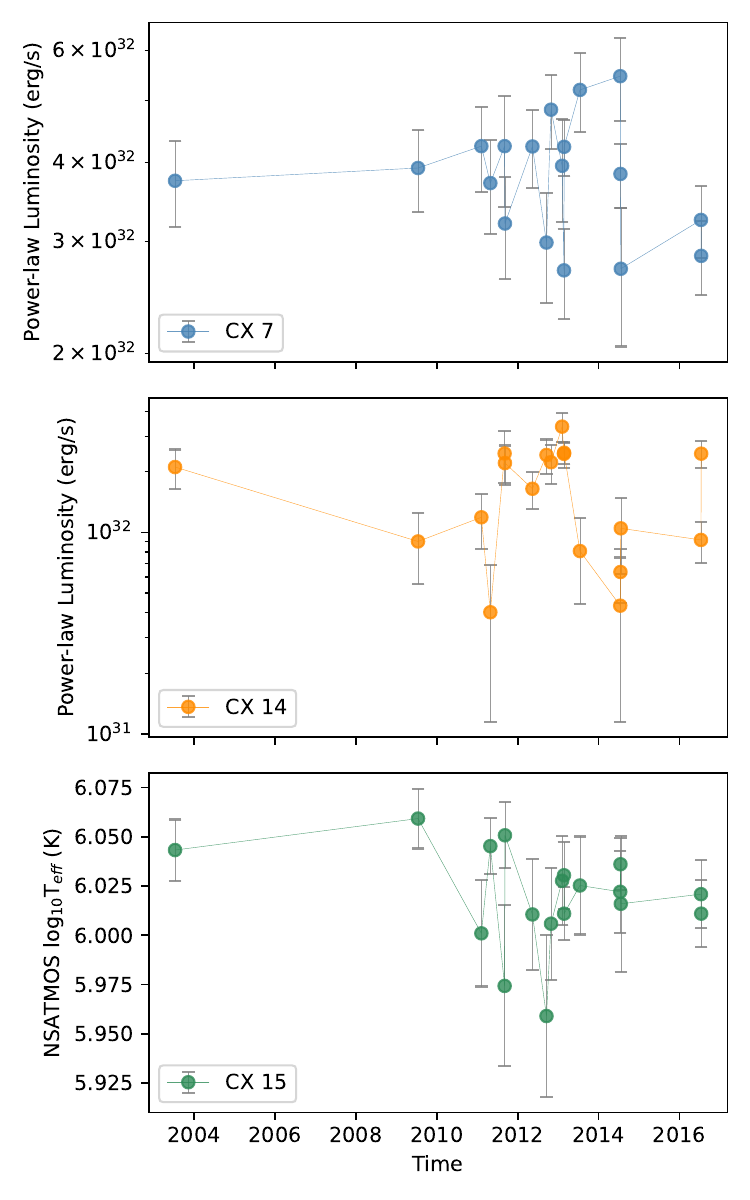}
    \caption{Variation of power-law luminosity (erg/s, calculated from flux normalization) for CX 7 and CX 14, and NSATMOS log$_{10}$T$_{eff}$ (K) for CX 15 as a function of time.}
    \label{fig:qlmxb_lc}
\end{figure}

We found that 10 of the 22 q/q? sources are variable. 
CX2 (Ter 5 X-3), CX3 (EXO 1745-248), and CX 25 (IGR J17480-2446) are known transient LMXBs, whose variability has been discussed elsewhere \citep{Degenaar15,RiveraSandoval2018}. 
CX 21 and CX 29 did not show long-term variability but exhibited variability within individual observations 14475 and 16638, respectively. We also attempted to determine whether the NSATMOS or power-law component contributes more to the variability in the sources exhibiting only long-term variability; CX 7, CX 14, CX 15, CX 23, and CX 56. For this analysis, we simultaneously fitted the spectra from all 18 individual observations with an NSATMOS+power-law model under one of two conditions: (i) keeping the NSATMOS effective temperature (\texttt{LogT\_eff}) free, or (ii) keeping the power-law \texttt{norm} free. The rest of the parameters were fixed at their best-fit values from the combined spectral fit. CX 7 and CX 14 are better fit by allowing the power-law flux normalization to vary, with final {\it wstat} fit statistics improving from 1439 to 1137 and from 822 to 812, respectively, varying the total $L_X$ by a factor of $>$2 on timescales less than a year. CX 15's variation must be attributed to the NSATMOS effective temperature, with fit statistics improving from 815 to 608, which we interpret as due to low-level accretion, as CX 15 did not undergo a major outburst during this time frame that would produce NS cooling. Figure \ref{fig:qlmxb_lc} illustrates the variation of the free parameter for these three sources over time. CX 23 and CX 56 did not show any significant difference in the final fit statistics between the two cases, likely due to insufficient counts for a robust analysis.  We also attempted to perform a simultaneous fit, keeping both \texttt{LogT\_eff} and \texttt{norm} free, but the final fit statistic did not improve by more than 1.0 compared to the best fit with only one free parameter.

\citet{Bahramian2015} searched for quiescent variability among the Terzan 5 qLMXBs CX9, CX12, CX 18, and CX 21, finding small long-term variations in CX 12 which they attribute to varying absorption, or variations in the  NSATMOS component. They also find short-term variation from CX21 in observation 14475, in agreement with our results. Our identification of variability from the NSATMOS component of CX 15 is the strongest signature yet seen of variation in the NSATMOS component, from a NS that has not been observed in outburst. The total accretion rate onto the NS was estimated to be \(4.5 \times 10^{-14} \, \text{M}_\odot\, \text{yr}^{-1}\) (\(2.8 \times 10^{12} \, \text{g}\, \text{s}^{-1}\)) by comparing the change in energy flux between the epochs of highest (ObsID: 10059) and lowest (ObsID: 14475) count rates for CX15.

\subsection{Interesting spectral feature in Ter 5 X-3}

\begin{figure*}
    \centering
    \includegraphics[scale=0.1]{ 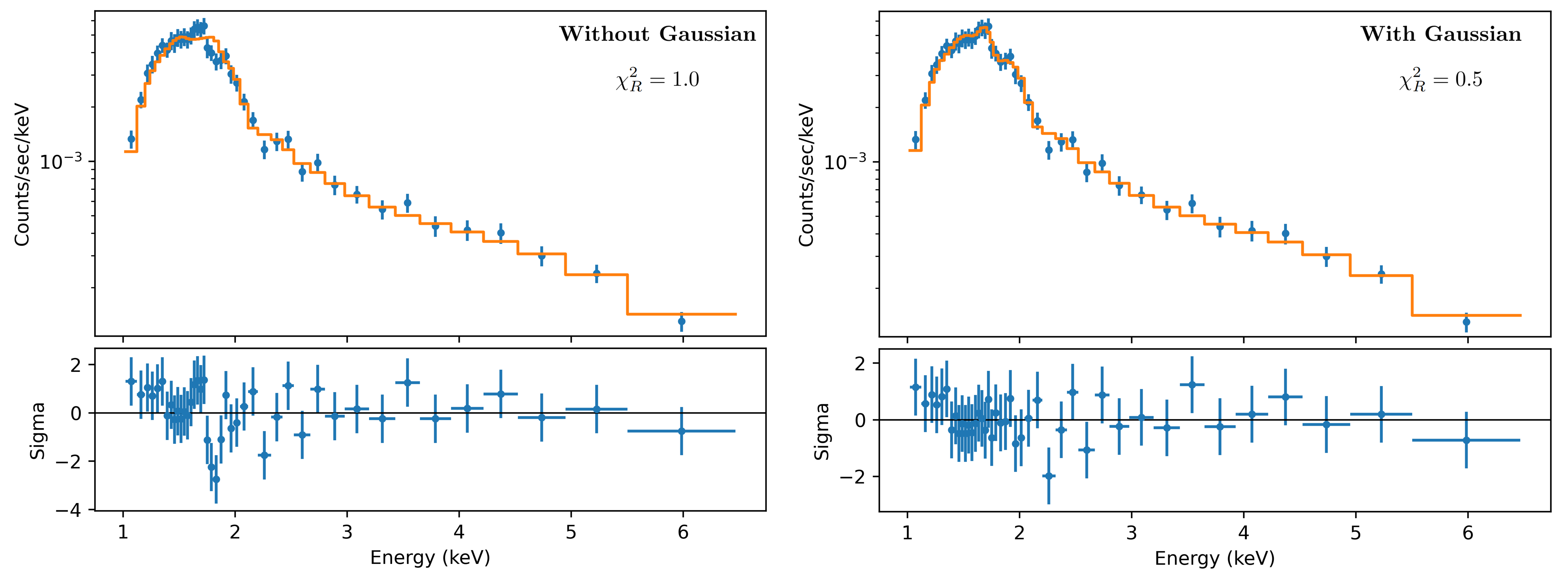}
    \caption{NSATMOS+power-law fit results and residuals for Ter 5 X-3 (CX 2). The left panel shows the fit without the Gaussian absorption and emission model, while the right panel includes it.}
    \label{fig:Ter5X3_spectra}
\end{figure*}

The combined spectrum of the qLMXB CX 2 (Ter 5 X-3) shows a fascinating residual around 1.7 keV (Fig. \ref{fig:Ter5X3_spectra}).
The residuals appear to show an emission feature and an absorption feature, as observed in P Cygni lines. 
We examined the individual observation spectra but could not identify any particular observation giving more weight to this feature; the feature appears to be present (but not significant) in all or most observations.  Other sources of similar brightness do not show similar features.

To improve the spectral fit, we added Gaussian emission and absorption lines to the NSATMOS+power-law fit using \texttt{SHERPA}'s one-dimensional Gaussian function (\texttt{gauss1d}). This reduced the reduced chi-square ($\chi_R^2$) from 1.0 to 0.5, as shown in Figure \ref{fig:Ter5X3_spectra}. The best-fit parameters for the Gaussian emission model were: \texttt{pos} (center of the Gaussian): 1.68 keV,   
 \texttt{fwhm}: 0.05,   
 \texttt{ampl} (maximum peak): \( 1.4 \times 10^{-5} \). For the Gaussian absorption model, the best-fit parameters were:
 \texttt{pos}: 1.8 keV,   
 \texttt{fwhm}: 0.05,   
 \texttt{ampl}: \( -1.9 \times 10^{-5} \). 
It would be hard to understand this as a real P Cygni feature from a wind, as the absorption line should be associated with a strong edge, as seen in Chandra grating spectroscopy \citep{Ueda05,Schulz16}. It is difficult to understand how this could be a feature on the NS surface, either, as elements such as Si should rapidly submerge under the hydrogen surface layer \citep{Alcock80}. An instrumental feature seems unlikely, but we tested whether a change in gain could improve the fit; it did not.

\cite{Bahramian2015} found an unusual spectral shape for CX 9, with a significant flux deficit at low energies (0.1–1 keV). They attributed this to either higher intrinsic absorption, possibly due to an edge-on viewing angle, or a helium-dominated atmosphere. However, in our analysis, we do not find any such spectral feature in the combined spectrum of CX 9. Additionally, we found that the He atmosphere model does not perform better than the H atmosphere model for any qLMXBs.  

\subsection{Power-law Sources} \label{subsec:pl_src}

We identified more than 30 X-ray sources whose spectra are best explained by a power-law. Figure \ref{fig:xray_cmd} shows that these sources exhibit hard X-ray spectra  (typically \texttt{PhoIndex}: 0–2), with $L_X$ ranging from \( 10^{31} \) to \( 10^{33} \, \text{erg/s} \). Fourteen of these power-law sources were found to be variable. CX 6 and CX 11 did not show long-term variability but exhibited variability within individual observations 15750 and 16638, respectively. 

Six power-law sources exhibit unusually hard spectra (\texttt{PhoIndex} \( \lesssim 1 \), see Table \ref{tab:pwrlw}). Three of these unusually hard sources are situated outside the core of Terzan 5, suggesting the possibility that they are background AGNs, showing additional intrinsic absorption. The remaining three unusually hard sources are located within the core radius of Terzan 5, suggesting they are cluster members. Among these, CX 10 is the MSP Terzan 5 P, with photon index 1.0$_{-0.1}^{+0.1}$, while CX 19, with photon index  1.1$_{-0.2}^{+0.2}$ is newly identified as the MSP Terzan 5 ar (see \S~7.5). The  remaining unusually hard core source, and the sources outside the core, could be CVs (such as intermediate polars which often have intrinsic absorption, or edge-on CVs) or MSPs. 

CX 53 (MSP at) also exhibits an unusually hard power-law spectrum with a fitted \texttt{PhoIndex} of $0.4^{+0.5}_{-0.6}$. However, this result is in discrepancy with its inferred \texttt{PhoIndex} from Figure~\ref{fig:xray_cmd}, which appears to be greater than 1. This source lies within 2$^{\prime\prime}$ of CX~3, the brightest X-ray source in the field, making it likely that the \texttt{ACIS\_EXTRACT} spectrum is contaminated by the nearby bright source (see Figure \ref{fig:XrayImage}) and affected by complications due to a non-local background selection. As a result, the derived \texttt{PhoIndex} from spectral fitting may be unreliable for this source, while the X-ray CMD (Figure \ref{fig:xray_cmd}) includes  local background subtraction, making the \texttt{PhoIndex} inferred from it more trustworthy. No other sources suffer such discrepancies.

\begin{table}
\caption{Spectral results of the power-law fit for the power-law sources.The derived photon index is denoted as \texttt{PhoIndex}. Note: Sources with fewer than 100 counts were fitted using \textit{wstat} statistics.}
    \centering
    \begin{tabular}{ccc}
    \toprule
       CX & \texttt{PhoIndex} & $\chi_R^2/dof$\\
        \toprule
1	&	1.4	$_{	-0.1}^{+0.1	}$&	0.75/16	\\
5	&	1.88	$_{-0.04}^{+0.04	}$&	0.77/37	\\
6	&	1.84	$_{	-0.07	}^{+	0.07	}$&	0.67/24	\\
8	&	1.6	$_{	-0.1	}^{+	0.1	}$&	1.04/14	\\
10	&	1.0	$_{	-0.1	}^{+	0.1	}$&	0.69/16	\\
11	&	1.54	$_{	-0.07	}^{+	0.07	}$&	0.74/18	\\
13	&	1.4	$_{	-0.1	}^{+	0.1	}$&	0.48/13	\\
16	&	1.7	$_{	-0.1	}^{+	0.1	}$&	1.2/11	\\
17	&	0.3	$_{	-0.2	}^{+	0.2	}$&	0.56/17	\\
19	&	1.1	$_{	-0.2	}^{+	0.2	}$&	0.47/19	\\
22	&	1.5	$_{	-0.1	}^{+	0.1	}$&	0.56/14	\\
24	&	1.9	$_{	-0.3	}^{+	0.3	}$&	0.21/14	\\
26	&	1.8	$_{	-0.2	}^{+	0.2	}$&	0.56/23	\\
28	&	1.1	$_{	-0.5	}^{+	0.5	}$&	0.38/13	\\
31	&	2.0	$_{	-0.2	}^{+	0.2	}$&	0.40/18	\\
33	&	1.8	$_{	-0.4	}^{+	0.4	}$&	1.60/9	\\
34	&	1.6	$_{	-0.2	}^{+	0.2	}$&	0.76/11	\\
37	&	1.2	$_{	-0.3	}^{+	0.2	}$&	0.91/14	\\
38	&	1.7	$_{	-0.1	}^{+	0.1	}$&	0.59/15	\\
41	&	1.5	$_{	-0.5	}^{+	0.5	}$&	1.03/5	\\
42	&	1.1	$_{	-0.4	}^{+	0.4	}$&	0.35/13	\\
43	&	2.0	$_{	-0.3	}^{+	0.3	}$&	0.78/19	\\
45	&	1.2	$_{	-0.2	}^{+	0.2	}$&	0.48/17	\\
46	&	1.6	$_{	-0.3	}^{+	0.3	}$&	0.52/15	\\
47	&	1.6	$_{	-0.2	}^{+	0.2	}$&	0.58/10	\\
48	&	0.7	$_{	-0.3	}^{+	0.3	}$&	0.72/10	\\
51	&	1.5	$_{	-0.7	}^{+	0.7	}$&	0.64/14	\\
53	&	0.4	$_{	-0.6	}^{+	0.5	}$&	0.50/15	\\
58	&	5.5	$_{	-0.8	}^{+	1.0	}$&	0.49/14	\\
59	&	2.9	$_{	-0.4	}^{+	0.4	}$&	0.43/9	\\
63	&	2.1	$_{	-0.6	}^{+	0.5	}$&	0.80/90	\\
72	&	1	$_{	-1	}^{+	1	}$&	0.68/26	\\
82	&	2.8	$_{	-0.5	}^{+	0.7	}$&	1.29/46	\\
83	&	3.2	$_{	-0.7	}^{+	0.7	}$&	0.69/43	\\
85	&	2.6	$_{	-0.3	}^{+	0.3	}$&	0.87/71	\\
98	&	2.7	$_{	-0.6	}^{+	0.6	}$&	0.98/21	\\ 
\hline 
    \end{tabular}
    \label{tab:pwrlw}
\end{table}

\subsection{Foreground and background sources} \label{sec:foreground&background}

The radial distribution calculation predicted of order 4 non-cluster sources within the half-mass radius, above 50 counts. We use equation 2 of \citet{Giacconi01}, with the cluster extinction, to predict 1.0 background AGN projected within the cluster radius (though $N_H$ may be higher through the full galaxy, leaving even fewer). This suggests that most non-cluster sources are foreground sources, which is in agreement with our identification of seven likely foreground sources by their X-ray colors (Fig.~\ref{fig:xraycmd_0.5-1}) and spectra (\S \ref{sec:spectralanalysis}). As the coronally active stars that dominate foreground sources tend to have low ratios of X-ray to visual flux ratio ($\text{Log}[f_x/f_v]$, \cite{Maccacaro1988}), we check for matches with bright stars.

We searched for counterparts using the \texttt{VizieR} Photometry viewer \footnote{\url{http://vizier.cds.unistra.fr/vizier/sed/beta/}}, with the search radius taken as 1.2 times the \textit{Chandra} 95\% confidence positional uncertainty (see Table \ref{tab:XraySources}). There are 374 Gaia sources with $G<19$ (not deep enough for Terzan 5's red giants) within Terzan 5's 0.72' half-mass radius, giving a density of 0.064 matches/arcsec$^2$. We then expect 0.05 spurious matches within our search areas. For CX 20 and CX 78, no clear optical counterpart was found within the search radius. We also searched in the infrared using \texttt{VizieR}, but did not find there counterparts.
But the other 5 candidate foreground sources have clear optical counterparts in Gaia, listed in Table~\ref{tab:foreground}. For CX 65 we converted the SDSS $g$-magnitude and $g - r$ color from \cite{Barry2008} 
using photometric transformations from \cite{Jordi2006}. \citet{BailerJones21} gives well-constrained parallax distances of 1.6$\pm0.1$ kpc for both CX 62 and CX 67, confirming them as foreground sources.  

Table \ref{tab:foreground} provides the logarithm of the X-ray to visual flux ratio ($\text{Log}[f_x/f_v]$, \cite{Maccacaro1988}) and the location of the source in the X-ray image. Comparing the $\text{Log}[f_x/f_v]$ values from Table \ref{tab:foreground} with Figure 1 from \cite{Maccacaro1988}, we find that the X-ray to visual flux ratios for CX 55, CX 62, CX 65, CX 66, and CX 67 agree with those of M/K-type stars, consistent with foreground stars. Spectral results from the APEC fit for likely foreground stars are also provided in Table \ref{tab:foreground}

\begin{table*}
\caption{Likely foreground sources. %identified in our analysis. 
Offset is the separation of the Gaia and Chandra source positions. The logarithm of the X-ray to visual flux ratio is given by $ \text{Log}[f_x/f_v] = log[f_x] + V/2.5 + 5.37 $ \citep{Maccacaro1988} Here, $f_x$ represents the X-ray flux in the 0.5–8 keV range. %and $m_V$ is the Johnson-Cousins apparent $V$-magnitude. %obtained from the \texttt{VizieR} Photometry Viewer. 
"Distance" %refers to the arcmin distance of the sources from 
is to 
the cluster's center. Spectral results from the APEC fit are also provided as $N_{H}$ (hydrogen column density), \texttt{kT} (plasma temperature), and energy flux (0.5--8 keV, from \texttt{calc\_energy\_flux}).
$^a$: \cite{Figuera2024}. 
$^b$: \cite{Girard2011}.
$^c$: \cite{Zacharias2012}.
$^d$: \cite{Barry2008}. 
 Note: Sources with fewer than 100 counts were fitted using \textit{wstat} statistics.
} %(R$_c$: 0.18$^\prime$, R$_h$: 0.72$^\prime$).}
\label{tab:foreground}
    \centering
    \begin{tabular}{ccccccccc}
        \toprule
        CX & Offset & V & Log[$f_x/f_v$] & Distance & $N_H$ & \texttt{kT} & Energy Flux & $\chi^2_R/dof$\\
         & (arcsec) &   &                & (arcmin)  & ($\times10^{22} cm^{-2}$) & (keV) & ($erg \,\,s^{-1} cm^{-2}$) & \\
        \toprule
20	& \nodata & \nodata &	\nodata & 0.40 & 0$^{+0.2}$		&	1.3	$^{+0.1}_{-0.2}$	&	1.29E-15	&	1.18/6	\\
55	& 0.14 & 17.7$^a$ &	-1.2 &	0.86  & 0.1$^{+0.3}_{-0.1}$	&	0.5$^{+0.2}_{-0.1}$	&	9.25E-16	&	0.43/9	\\	
62	& 0.10 & 16.5$^b$ &	-2.1 &	0.30 & 1.0$^{+0.2}_{-0.2}$	&	0.5$^{+0.1}_{-0.1}$	&	6.51E-16	&	0.62/54	\\ 		
65	& 0.054 & 19.4$^d$ &	-0.9 &	0.84 & 1.4$^{+0.3}_{-0.2}$	&	0.7$^{+0.3}_{-0.2}$	&	7.12E-16	&	0.99/67	\\ 		
66	& 0.065 & 16.5$^b$ &	-2.1 &	0.31 & 0$^{+0.1}$			&	1.3$^{+0.2}_{-0.2}$	&	6.14E-16	&	1.29/47	\\		
67	& 0.026 & 17.4$^c$ &	-1.8 &	0.60 & 1.0$^{+0.2}_{-0.2}$	&	0.7$_{-0.1}^{+0.1}$	&	5.75E-16	&	1.00/63	\\		
78	& \nodata & \nodata &	\nodata  &	 0.60 & 0.9$^{+0.2}_{-0.3}$	&	0.6$_{-0.1}^{+0.3}$	&	3.81E-16	&	1.08/41	\\	
        \hline 
    \end{tabular}
\end{table*}

\subsection{MSPs}

Terzan 5 hosts the largest MSP population among all the known GCs, likely due to its exceptionally high stellar density and stellar encounter rate. To date, 49 MSPs have been discovered in Terzan 5 \citep[e.g.,][]{Ransom2005,Prager2017,Padmanabh2024}, while population estimates suggest that 100--200 MSPs may reside in the cluster based on various modeling approaches \citep[e.g.,][]{Bagchi2011,Zhao2022,Yin2024}. \citet{Bogdanov2021} searched for X-ray counterparts to MSPs Terzan 5 A through Terzan 5 an. In this section, we extend this search to MSPs Terzan 5 ao through ax, which have new timing solutions \citep{Padmanabh2024}. For Terzan 5 ar we adopted the timing information from \citet{Corcoran2024}.

We identified three X-ray sources (CX 29, CX 19, and CX 53) in close proximity to MSPs Terzan 5 aq, ar, and at, with angular separations of 0.29", 0.21", and 0.10", respectively. Terzan 5 aq is a BW pulsar, possibly showing radio eclipses and having a minimum companion mass of $\sim$0.013 M$_\odot$ \citep{Padmanabh2024}. Eclipsing BWs typically exhibit X-ray luminosities of $\sim10^{31}~{\rm erg~s}^{-1}$, dominated by hard nonthermal emission \citep{Zhao2022}. However, CX29, the adjacent X-ray source to Terzan 5 aq, has an X-ray luminosity of $(4.27\pm0.36) \times 10^{32}~{\rm erg~s}^{-1}$, with almost all its 
X-rays described as thermal emission from a neutron star surface (see, e.g., Figure~\ref{fig:xray_cmd}), suggesting this X-ray source is a qLMXB. Also, the 95\% confidence positional uncertainty of CX 29 is 0.13", much smaller than the angular separation (0.29") to the timing position of Terzan 5 aq. Therefore, we suggest that CX 29 is unlikely to be the X-ray counterpart to Terzan 5 aq. 

CX 19 and its radio counterpart Ter5-VLA38 had been previously proposed to be a RB pulsar, based on its X-ray spectral properties, X-ray orbital modulation, and radio spectral index \citep{Urquhart2020}. The subsequent radio detection of Terzan 5 ar near CX 19's position, along with the matching orbital period of 0.5133 days in both X-ray and radio observations, strongly supports that CX 19 is indeed the X-ray counterpart to Terzan 5 ar \citep{Padmanabh2024}. Although the timing position of Terzan 5 ar lies outside the 95\% confidence positional uncertainty of CX 19, the phase connection could be maintained when 
the timing position was fixed to CX 19's source coordinates, confirming their link \citep{Padmanabh2024}. Considering Terzan 5 ar is a RB, the positional offset between its timing and X-ray positions is likely due to its significant orbital variability leading to timing solution difficulties \citep[see e.g.,][]{Padmanabh2024,Corcoran2024,Rosenthal2025}.

\begin{figure}
    \centering
    \includegraphics[width=\linewidth]{ 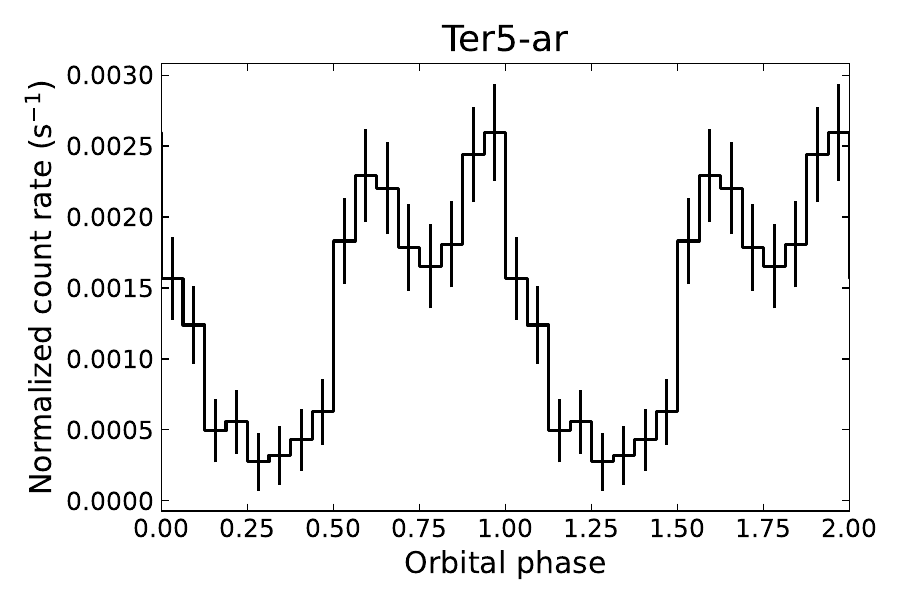}
    \includegraphics[width=\linewidth]{ 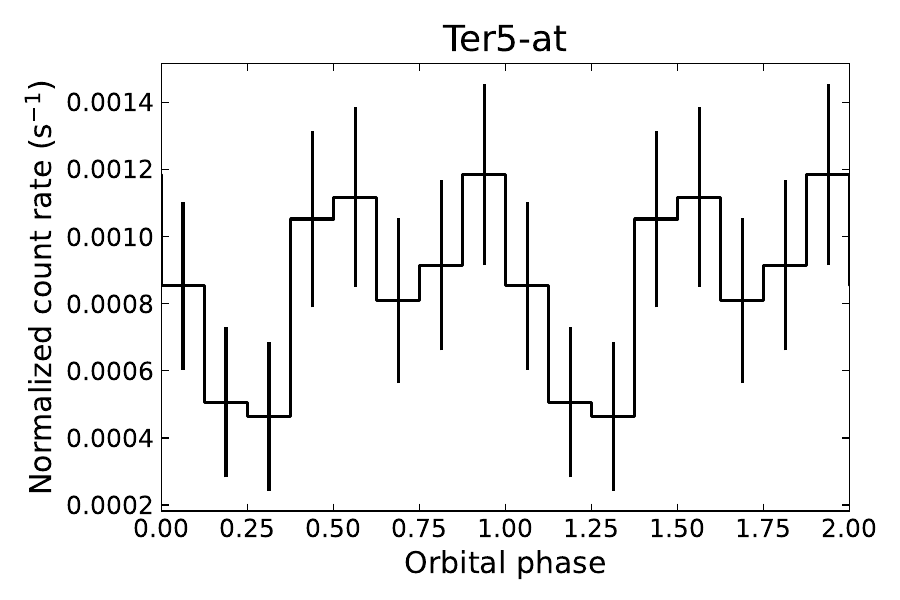}
    \caption{The background-subtracted, binned light curves of Terzan 5 ar (upper) and Terzan 5 at (lower), observed by {\it Chandra} ACIS-S in 0.5--8 keV. The light curves are folded at the orbital periods of 0.5133 and 0.2189 days, %for MSP ar and at, 
    respectively \citep{Padmanabh2024}, while the orbital phase zero is defined as the pulsar passage through the ascending node. Two orbital cycles are shown for clarity.}
    \label{fig:ter5-ar}
\end{figure}

We incorporate Terzan 5 ar's timing information to produce a more accurate phase-resolved light curve (Figure~\ref{fig:ter5-ar}). The phased light curve shows a distinct double-peaked morphology, similar to that observed in other RB systems such as PSR J2129$-$0429 \citep{AlNoori2018} and PSR J2339$-$0533 \citep{Kandel2019}. This configuration is consistent with intrabinary shock emission, particularly in systems where a strong stellar wind from the companion causes the shock to wrap around the pulsar \citep{Romani2016,Wadiasingh2017,Kandel2019}. Further studies of Terzan 5 ar, coupled with optical observations, may place strong constraints on the properties and geometry of this system \citep[e.g.][]{Sullivan2024}.

Terzan 5 at is an eclipsing BW with a minimum companion mass of 0.035 M$_\odot$ and an orbital period of $\sim$0.22 days \citep{Padmanabh2024}. The potential X-ray counterpart, CX 53, has an angular separation of only 0.10" from MSP at's timing position, within the 95\% confidence positional uncertainty (0.13") of CX 53, which significantly reduces the chance of coincidence. Furthermore, CX 53 shows a hard power-law spectrum with a photon index inferred from the CMD of approximately 1.2 (see Section~\ref{subsec:pl_src}), consistent with typical eclipsing BWs \citep[e.g.][]{Bogdanov2006}. %Additionally, based on 
Using the correlation between spider pulsar minimum companion mass and X-ray luminosity of \citet{Zhao2023}, the X-ray luminosity (0.5--10 keV) of Terzan 5 can be estimated as  $1.1^{+2.8}_{-0.8} \times 10^{31}$ erg~s$^{-1}$, consistent with the X-ray luminosity of CX 53. Hence, we suggest that CX 53 is likely the X-ray counterpart to MSP at, given its position, X-ray luminosity and spectrum. No variability was found from CX 53, likely owing to its relatively faint nature. 
Nevertheless, we created the phase-folded X-ray light curve for CX 53 using the timing solution of MSP at (lower panel of Figure~\ref{fig:ter5-ar}). Its light curve shows a suggestion of orbital phase-dependent variability, consistent with the expected phasing (as for Terzan 5 ar). By running Kuiper tests with 10,000 Monte Carlo resamplings, we found 97\% of the tests have $p$-values less than 0.05, strongly rejecting the null hypothesis that the observed modulation is distributed according to a uniform distribution. The observed light curve can also be explained as intrabinary shock emission by manipulating the model parameters, such as the wind momentum ratio and inclination angle \citep[see e.g.][]{Romani2016,Wadiasingh2017}. This brings the total number of spider pulsars in Terzan 5 with observed orbital X-ray modulations to five.

\section{Conclusion}

We have studied the photometry, spectra, and variability of the faint X-ray sources in the globular cluster Terzan 5, using 737 ks of archival {\it Chandra} data. Creating X-ray color-magnitude diagrams using two different X-ray colors, we clearly see three groups of sources; foreground sources (which have substantial emission below 1 keV, as they are not strongly absorbed like cluster sources), qLMXBs, and other cluster sources. These distinctions are confirmed by spectral fitting, and by identifying optical counterparts for 5 of the 7 putative foreground sources. 

We identify 22 candidate qLMXBs (7 of which are not certain) in Terzan 5, more than observed in any other globular cluster. 
The qLMXBs are well described by a hydrogen neutron star atmosphere plus a power-law, with the power-law usually making up a smaller fraction of the total flux. The transient LMXB CX 2 (or Ter 5 X-3) shows an odd residual in its spectra around 1.8 keV, which can be fit by emission and absorption lines. We do not see qLMXBs dominated by thermal emission below $L_X\sim10^{32}$ erg/s; if they are present, their power-law component must dominate their spectra. We see variability in 10 of the 22 qLMXBs. Some of this variation is due to varying power-law fluxes, but the variation in CX 15 seems to be variation of the NS temperature.  

The distribution of qLMXBs, and of other X-ray sources (above a 50-count limit), within the cluster is more concentrated than that of stars around the main-sequence turnoff. This is consistent with expectations for relaxed stellar populations with higher mass than the reference population; we infer an average mass of 1.46$\pm0.14$ \msun\ for the qLMXBs, and $1.34\pm0.13$ \msun\ for the non-qLMXB population. 

We verify association of the newly discovered MSP Terzan 5 ar with the hard X-ray source CX 19 by phase-folding the X-ray light curve to reveal the characteristic double-peaked curve of intrabinary shock emission, as seen in other redback MSPs. We identify a new association of the eclipsing black widow MSP Terzan 5 at with CX 53, due to the match of positions, CX 53's X-ray luminosity and spectral hardness matching expectations, and its phased X-ray light curve showing a double-peaked signature of intrabinary shock emission as well. 

\section*{acknowledgments}

CH is supported by NSERC Discovery Grant RGPIN-2023-04264, and Alberta Innovates Advance Program 242506334. 
JZ is supported by China Scholarship Council (CSC), File No. 202108180023. 

This paper employs a list of Chandra datasets, obtained by the Chandra X-ray Observatory, contained in the Chandra Data Collection (CDC) 387~\dataset[doi:10.25574/cdc.387]{https://doi.org/10.25574/cdc.387}.

This research has made use of data obtained from the Chandra Data Archive, and software provided by the Chandra X-ray Center (CXC) in the application packages \texttt{CIAO} \citep{Fruscione2006}, \texttt{SHERPA} \citep{Freeman2001, Doe2007, Burke2023}, and \texttt{DS9} \citep{SAO2000, SAO2003}. 

This research has made use of the VizieR catalogue access tool, CDS, Strasbourg, France (DOI : 10.26093/cds/vizier). The original description of the VizieR service was published in 2000, A\&AS 143, 23. 

\bibliography{references}{}
\bibliographystyle{aasjournal}



\appendix

\renewcommand{\thetable}{A\arabic{table}}
\setcounter{table}{0} 

Table \ref{tab:XraySources} provides information on the 126 X-ray sources detected within 2' x 2' of the cluster center in the merged
\textit{Chandra} ACIS-S 0.5-8 keV image of Terzan 5.

\begin{longtable}{ccccccccc}
    \caption{X-ray Sources in Terzan 5. CXOU\_J identifiers follow the Chandra convention for naming unregistered sources. The CXOU identifiers marked with an ``*”  are the sources not detected in \cite{Bahramian2020}. The right ascension (R.A.), declination (Decl.), and counts in 0.5–8 keV are provided by the \texttt{wavdetect} algorithm. The R.A. and Decl. are corrected for the radio positions, as discussed in \S~\ref{sec:sourcedet}. PU are the \textit{Chandra} 95\% confidence positional uncertainty in arcsec, calculated from equation 12 of \cite{Kim2007}. The methodology to derive unabsorbed luminosities, $L_X$ ($10^{32}$ erg s$^{-1}$), and the hardness ratio, $X_C = 2.5 \log_{10}(\text{cts}_{0.5-2.0} / \text{cts}_{2.0-8.0})$, is given in \S~\ref{sec:xraycmd}. Notes indicate qLMXB (q), maybe qLMXB (q?), powerlaw (p), foreground (f) and variable (V). } \\
     \toprule 
    CX  & CXOU\_J & R.A. & Decl.& PU & Counts  & L$_X$ & X$_C$ & NOTES 
   \\
     & &  &  & (arcsec) & (0.5-8 keV) & ($10^{32}$ erg s$^{-1}$) &  &  
   \\
   \toprule 
    \endfirsthead
\multicolumn{9}{c}{{\textit{Continued on next page}}} \\
\endfoot
\hline
\endlastfoot

1	&	174804.58-244641.8	&	267.01907	&	-24.77840	&	0.08	&	1737	$\pm$	45	&	4.71	$\pm$	0.19	&	-1.39	$\pm$	0.07	&	p, V	\\
2	&	174805.41-244637.6	&	267.02253	&	-24.77721	&	0.07	&	4660	$\pm$	70	&	23	$\pm$	0.35	&	0.50	$\pm$	0.03	&	q, V, Ter 5 X-3	\\
3	&	174805.23-244647.3	&	267.02176	&	-24.77991	&	0.06	&	10596	$\pm$	106	&	36	$\pm$	0.36	&	-0.58	$\pm$	0.02	&	q, V, EXO1745-248	\\
4	&	174804.71-244709.0	&	267.01960	&	-24.78593	&	0.07	&	4010	$\pm$	64	&	12.5	$\pm$	0.2	&	-0.82	$\pm$	0.04	&	q	\\
5	&	174802.66-244602.4	&	267.01107	&	-24.76744	&	0.07	&	4306	$\pm$	67	&	10.9	$\pm$	0.25	&	-0.95	$\pm$	0.04	&	p	\\
6	&	174804.42-244638.1	&	267.01840	&	-24.77735	&	0.07	&	2772	$\pm$	55	&	6.95	$\pm$	0.24	&	-1.00	$\pm$	0.05	&	p, V	\\
7	&	174804.11-244640.3	&	267.01709	&	-24.77798	&	0.08	&	2224	$\pm$	49	&	7.50	$\pm$	0.18	&	-1.32	$\pm$	0.06	&	q?, V	\\
8	&	174804.40-244703.6	&	267.01829	&	-24.78444	&	0.09	&	854	$\pm$	30	&	2.01	$\pm$	0.1	&	-1.18	$\pm$	0.09	&	p, V	\\
9	&	174804.82-244644.7	&	267.02008	&	-24.77921	&	0.08	&	1525	$\pm$	43	&	10.8	$\pm$	0.31	&	1.01	$\pm$	0.07	&	q	\\
10	&	174805.05-244641.0	&	267.02099	&	-24.77815	&	0.08	&	1766	$\pm$	45	&	4.32	$\pm$	0.16	&	-1.84	$\pm$	0.08	&	p, V, MSP P	\\
11	&	174804.25-244642.2	&	267.01766	&	-24.77848	&	0.08	&	2015	$\pm$	47	&	4.96	$\pm$	0.15	&	-1.31	$\pm$	0.06	&	p, V	\\
12	&	174806.20-244642.6	&	267.02582	&	-24.77860	&	0.08	&	1958	$\pm$	45	&	10.3	$\pm$	0.24	&	0.19	$\pm$	0.05	&	q	\\
13	&	174803.86-244641.5	&	267.01604	&	-24.77830	&	0.09	&	798	$\pm$	30	&	1.84	$\pm$	0.1	&	-1.28	$\pm$	0.10	&	p, V, MSP ad	\\
14	&	174805.38-244652.6	&	267.02238	&	-24.78141	&	0.09	&	918	$\pm$	32	&	4.8	$\pm$	0.17	&	-0.28	$\pm$	0.08	&	q, V	\\
15	&	174804.20-244647.9	&	267.01746	&	-24.78008	&	0.09	&	873	$\pm$	32	&	6.22	$\pm$	0.23	&	1.34	$\pm$	0.10	&	q, V	\\
16	&	174803.58-244649.4	&	267.01490	&	-24.78049	&	0.10	&	699	$\pm$	28	&	1.74	$\pm$	0.11	&	-1.09	$\pm$	0.09	&	p, V	\\
17	&	174804.34-244635.9	&	267.01807	&	-24.77676	&	0.09	&	993	$\pm$	34	&	2.62	$\pm$	0.15	&	-2.35	$\pm$	0.12	&	p	\\
18	&	174805.28-244651.3	&	267.02196	&	-24.78103	&	0.09	&	1153	$\pm$	36	&	17.2	$\pm$	0.54	&	0.95	$\pm$	0.08	&	q	\\
19	&	174804.63-244645.2	&	267.01925	&	-24.77935	&	0.09	&	880	$\pm$	34	&	2.34	$\pm$	0.16	&	-1.82	$\pm$	0.12	&	p, V, MSP ar	\\
20	&	174803.07-244640.7	&	267.01277	&	-24.77809	&	0.13	&	199	$\pm$	15	&	0.97	$\pm$	0	&	2.19	$\pm$	0.25	&	f, V	\\
21	&	174804.28-244625.4	&	267.01780	&	-24.77383	&	0.09	&	910	$\pm$	31	&	6.66	$\pm$	0.23	&	1.38	$\pm$	0.09	&	q, V	\\
22	&	174806.19-244617.6	&	267.02573	&	-24.77164	&	0.09	&	835	$\pm$	30	&	1.98	$\pm$	0.09	&	-1.26	$\pm$	0.09	&	p	\\
23	&	174803.53-244645.9	&	267.01467	&	-24.77950	&	0.13	&	182	$\pm$	15	&	1.03	$\pm$	0.06	&	-0.93	$\pm$	0.19	&	q?, V	\\
24	&	174805.12-244646.0	&	267.02129	&	-24.77957	&	0.11	&	324	$\pm$	21	&	0.92	$\pm$	0.15	&	-1.18	$\pm$	0.17	&	p, V	\\
25	&	174804.83-244648.7	&	267.02010	&	-24.78031	&	0.08	&	1447	$\pm$	41	&	19.7	$\pm$	0.56	&	1.66	$\pm$	0.09	&	q, V, IGR J17480-2446	\\
26	&	174803.86-244645.9	&	267.01607	&	-24.77952	&	0.11	&	447	$\pm$	23	&	1.18	$\pm$	0.1	&	-1.08	$\pm$	0.12	&	p	\\
27	&	174806.10-244624.1	&	267.02539	&	-24.77345	&	0.10	&	537	$\pm$	24	&	9.35	$\pm$	0.42	&	2.02	$\pm$	0.15	&	q	\\
28	&	174804.69-244648.4	&	267.01949	&	-24.78020	&	0.12	&	287	$\pm$	21	&	0.59	$\pm$	0.1	&	-1.91	$\pm$	0.25	&	p	\\
29	&	174804.75-244642.5	&	267.01979	&	-24.77861	&	0.13	&	200	$\pm$	17	&	4.27	$\pm$	0.36	&	3.20	$\pm$	2.50	&	q, V	\\
30	&	174804.58-244640.5	&	267.01908	&	-24.77800	&	0.11	&	354	$\pm$	22	&	3.71	$\pm$	0.23	&	3.12	$\pm$	1.72	&	q	\\
31	&	174804.24-244700.7	&	267.01764	&	-24.78362	&	0.11	&	406	$\pm$	21	&	1.1	$\pm$	0.09	&	-0.84	$\pm$	0.12	&	p	\\
32	&	174805.36-244631.5	&	267.02231	&	-24.77553	&	0.11	&	385	$\pm$	21	&	1.93	$\pm$	0.14	&	-0.64	$\pm$	0.12	&	q?	\\
33	&	174804.78-244650.7	&	267.01987	&	-24.78087	&	0.12	&	208	$\pm$	18	&	0.42	$\pm$	0.08	&	-1.19	$\pm$	0.22	&	p	\\
34	&	174804.70-244604.9	&	267.01954	&	-24.76812	&	0.12	&	275	$\pm$	17	&	0.64	$\pm$	0.06	&	-1.21	$\pm$	0.15	&	p	\\
35	&	174805.01-244652.8	&	267.02082	&	-24.78143	&	0.14	&	127	$\pm$	14	&	0.41	$\pm$	0.05	&	-0.92	$\pm$	0.25	&	\nodata	\\
36	&	174805.70-244642.9	&	267.02368	&	-24.77867	&	0.13	&	152	$\pm$	13	&	0.97	$\pm$	0.35	&	-0.38	$\pm$	0.20	&	q?	\\
37	&	174804.62-244652.3	&	267.01919	&	-24.78130	&	0.11	&	312	$\pm$	20	&	0.68	$\pm$	0.07	&	-2.14	$\pm$	0.24	&	p, V	\\
38	&	174805.38-244656.2	&	267.02241	&	-24.78239	&	0.09	&	881	$\pm$	31	&	2.17	$\pm$	0.11	&	-1.17	$\pm$	0.09	&	p, V	\\
39	&	174804.92-244642.7	&	267.02051	&	-24.77862	&	0.18	&	78	$\pm$	13	&	0.25	$\pm$	0.04	&	-1.96	$\pm$	1.69	&	\nodata	\\
40	&	174804.65-244625.0	&	267.01934	&	-24.77375	&	0.17	&	92	$\pm$	10	&	0.3	$\pm$	0.03	&	-2.07	$\pm$	0.39	&	\nodata	\\
41	&	174804.23-244624.3	&	267.01758	&	-24.77350	&	0.12	&	270	$\pm$	17	&	0.54	$\pm$	0.09	&	-1.87	$\pm$	0.61	&	p, V	\\
42	&	174804.89-244628.3	&	267.02032	&	-24.77464	&	0.14	&	129	$\pm$	12	&	0.3	$\pm$	0.04	&	-1.56	$\pm$	0.27	&	p	\\
43	&	174804.01-244647.2	&	267.01667	&	-24.77986	&	0.13	&	194	$\pm$	16	&	0.43	$\pm$	0.09	&	-1.25	$\pm$	0.21	&	p	\\
44	&	174804.43-244632.8	&	267.01843	&	-24.77591	&	0.21	&	59	$\pm$	9	&	0.19	$\pm$	0.03	&	-0.94	$\pm$	0.34	&	\nodata	\\
45	&	174805.27-244640.0	&	267.02193	&	-24.77787	&	0.12	&	306	$\pm$	20	&	0.74	$\pm$	0.06	&	-1.47	$\pm$	0.17	&	p, V	\\
46	&	174807.43-244658.3	&	267.03092	&	-24.78297	&	0.13	&	166	$\pm$	13	&	0.38	$\pm$	0.05	&	-1.26	$\pm$	0.19	&	p, V	\\
47	&	174804.24-244606.6	&	267.01761	&	-24.76862	&	0.12	&	246	$\pm$	16	&	0.61	$\pm$	0.05	&	-1.08	$\pm$	0.15	&	p	\\
48	&	174806.33-244637.5	&	267.02634	&	-24.77719	&	0.12	&	257	$\pm$	17	&	0.56	$\pm$	0.05	&	-1.95	$\pm$	0.22	&	p	\\
\footnote{We did not find any counterpart to CX 49 \citep{Heinke2006} in our analysis. There is no detection at its position in the combined dataset. It is only detected in the 2003 observation (Observation ID: 3798), suggesting that it is plausibly a transient object that was active only during that epoch.}50	&	174805.89-244646.4	&	267.02451	&	-24.77964	&	0.23	&	46	$\pm$	8	&	1.16	$\pm$	0.04	&	2.89	$\pm$	3.62	&	q?	\\
51	&	174805.06-244645.7	&	267.02108	&	-24.77945	&	0.13	&	168	$\pm$	16	&	0.31	$\pm$	0.09	&	-1.19	$\pm$	0.55	&	p	\\
52	&	174804.60-244648.4	&	267.01915	&	-24.78020	&	0.13	&	158	$\pm$	15	&	1.13	$\pm$	0.15	&	-0.48	$\pm$	0.22	&	q?	\\
53	&	174805.37-244646.7	&	267.02234	&	-24.77977	&	0.13	&	152	$\pm$	15	&	0.33	$\pm$	0.07	&	-1.42	$\pm$	0.25	&	p, MSP at	\\
54	&	174805.66-244642.0	&	267.02354	&	-24.77841	&	0.13	&	146	$\pm$	13	&	2.84	$\pm$	0.25	&	1.71	$\pm$	0.79	&	q	\\
55	&	174801.52-244619.5	&	267.00630	&	-24.77219	&	0.14	&	137	$\pm$	12	&	1.15	$\pm$	0.1	&	2.27	$\pm$	0.29	&	f	\\
56	&	174804.80-244633.6	&	267.01999	&	-24.77610	&	0.14	&	137	$\pm$	14	&	0.93	$\pm$	0.09	&	0.26	$\pm$	0.21	&	q, V	\\
57	&	174803.93-244638.1	&	267.01636	&	-24.77737	&	0.14	&	129	$\pm$	14	&	0.42	$\pm$	0.05	&	-1.50	$\pm$	0.27	&	V	\\
58	&	174804.84-244639.1	&	267.02014	&	-24.77761	&	0.15	&	109	$\pm$	13	&	0.53	$\pm$	0.06	&	0.82	$\pm$	0.28	&	p	\\
59	&	174808.74-244648.5	&	267.03639	&	-24.78023	&	0.15	&	109	$\pm$	11	&	0.53	$\pm$	0.05	&	0.12	$\pm$	0.22	&	p	\\
60	&	174804.44-244543.7$^*$	&  	267.01850	&	-24.76213	&	0.16	&	101	$\pm$	10	&	0.33	$\pm$	0.03	&	-1.02	$\pm$	0.23	&	\nodata	\\
61	&	174804.61-244634.2	&	267.01917	&	-24.77630	&	0.16	&	96	$\pm$	11	&	0.31	$\pm$	0.04	&	-2.10	$\pm$	0.38	&	\nodata	\\
62	&	174805.83-244655.7	&	267.02423	&	-24.78224	&	0.17	&	92	$\pm$	11	&	0.45	$\pm$	0.05	&	2.71	$\pm$	2.17	&	f	\\
63	&	174804.69-244651.1	&	267.01954	&	-24.78099	&	0.17	&	89	$\pm$	12	&	0.29	$\pm$	0.04	&	-0.33	$\pm$	0.28	&	MSP O	\\
64	&	174808.85-244630.5	&	267.03684	&	-24.77522	&	0.17	&	88	$\pm$	10	&	0.29	$\pm$	0.03	&	-0.12	$\pm$	0.24	&	\nodata	\\
65	&	174801.52-244621.3	&	267.00631	&	-24.77269	&	0.17	&	88	$\pm$	10	&	0.43	$\pm$	0.05	&	1.49	$\pm$	0.34	&	f	\\
66	&	174803.54-244650.9	&	267.01472	&	-24.78092	&	0.17	&	84	$\pm$	10	&	0.41	$\pm$	0.05	&	2.78	$\pm$	2.54	&	f	\\
67	&	174802.38-244658.0	&	267.00990	&	-24.78285	&	0.18	&	80	$\pm$	9	&	0.39	$\pm$	0.04	&	1.52	$\pm$	0.30	&	f	\\
68	&	174809.30-244737.6$^*$	&	267.03877	&	-24.79378	&	0.19	&	68	$\pm$	9	&	0.22	$\pm$	0.03	&	-1.16	$\pm$	0.31	&	\nodata	\\
69	&	174804.71-244646.9	&	267.01963	&	-24.77980	&	0.19	&	68	$\pm$	12	&	0.22	$\pm$	0.04	&	-0.20	$\pm$	0.35	&	\nodata	\\
70	&	174804.57-244650.3	&	267.01901	&	-24.78075	&	0.19	&	67	$\pm$	11	&	0.22	$\pm$	0.04	&	-1.69	$\pm$	0.49	&	\nodata	\\
71	&	174804.30-244627.3	&	267.01783	&	-24.77437	&	0.21	&	58	$\pm$	9	&	0.19	$\pm$	0.03	&	-2.16	$\pm$	1.87	&	\nodata	\\
72	&	174802.26-244637.5	&	267.00938	&	-24.77716	&	0.21	&	55	$\pm$	8	&	0.18	$\pm$	0.03	&	-1.19	$\pm$	0.39	&	V, MSP A	\\
73	&	174805.40-244644.9	&	267.02247	&	-24.77925	&	0.21	&	55	$\pm$	10	&	0.18	$\pm$	0.03	&	-1.92	$\pm$	1.78	&	\nodata	\\
74	&	174804.08-244645.4	&	267.01693	&	-24.77937	&	0.21	&	55	$\pm$	9	&	0.18	$\pm$	0.03	&	-2.01	$\pm$	0.58	&	V	\\
75	&	174805.44-244624.1	&	267.02262	&	-24.77349	&	0.21	&	55	$\pm$	8	&	0.18	$\pm$	0.03	&	-0.66	$\pm$	0.34	&	\nodata	\\
76	&	174803.77-244642.4	&	267.01566	&	-24.77854	&	0.22	&	53	$\pm$	9	&	0.17	$\pm$	0.03	&	-0.69	$\pm$	0.36	&	V	\\
77	&	174804.54-244654.6	&	267.01891	&	-24.78195	&	0.22	&	52	$\pm$	9	&	0.17	$\pm$	0.03	&	-0.92	$\pm$	0.42	&	\nodata	\\
78	&	174803.08-244617.8	&	267.01279	&	-24.77172	&	0.23	&	48	$\pm$	7	&	0.23	$\pm$	0.03	&	1.32	$\pm$	1.00	&	f, V	\\
79	&	174801.85-244722.2	&	267.00768	&	-24.78958	&	0.24	&	44	$\pm$	7	&	0.14	$\pm$	0.02	&	-1.75	$\pm$	0.58	&	\nodata	\\
80	&	174804.20-244658.6	&	267.01745	&	-24.78302	&	0.24	&	43	$\pm$	8	&	0.14	$\pm$	0.03	&	-1.60	$\pm$	1.37	&	\nodata	\\
81	&	174801.07-244603.3$^*$	&	267.00444	&	-24.76757	&	0.24	&	43	$\pm$	7	&	0.21	$\pm$	0.03	&	1.78	$\pm$	1.41	&	\nodata	\\
82	&	174803.31-244632.2	&	267.01382	&	-24.77569	&	0.25	&	40	$\pm$	7	&	0.2	$\pm$	0.03	&	0.19	$\pm$	0.39	&	p	\\
83	&	174805.10-244634.2	&	267.02123	&	-24.77626	&	0.25	&	40	$\pm$	8	&	0.13	$\pm$	0.03	&	0.00	$\pm$	0.61	&	MSP V	\\
84	&	174804.23-244703.1	&	267.01760	&	-24.78427	&	0.25	&	39	$\pm$	7	&	0.13	$\pm$	0.02	&	-1.16	$\pm$	1.11	&	\nodata	\\
85	&	174804.42-244641.3	&	267.01834	&	-24.77828	&	0.26	&	38	$\pm$	9	&	0.19	$\pm$	0.04	&	0.37	$\pm$	0.50	&	p	\\
86	&	174803.18-244638.0	&	267.01323	&	-24.77731	&	0.28	&	33	$\pm$	7	&	0.11	$\pm$	0.02	&	-1.19	$\pm$	0.47	&	\nodata	\\
87	&	174806.25-244546.4$^*$               	&	267.02603	&	-24.76290	&	0.28	&	31	$\pm$	6	&	0.1	$\pm$	0.02	&	-1.25	$\pm$	0.61	&	\nodata	\\
88	&	174809.31-244634.6$^*$	&	267.03879	&	-24.77627	&	0.29	&	30	$\pm$	6	&	0.1	$\pm$	0.02	&	-2.39	$\pm$	2.90	&	\nodata	\\
89	&	174807.43-244602.6	&	267.03091	&	-24.76749	&	0.29	&	30	$\pm$	6	&	0.1	$\pm$	0.02	&	-1.75	$\pm$	1.75	&	\nodata	\\
90	&	174805.70-244639.8	&	267.02372	&	-24.77782	&	0.29	&	30	$\pm$	7	&	0.1	$\pm$	0.02	&	-0.14	$\pm$	0.78	&	\nodata	\\
91	&	174804.35-244645.6	&	267.01808	&	-24.77943	&	0.26	&	38	$\pm$	7	&	1.22	$\pm$	0.03	&	2.67	$\pm$	3.99	&	q?	\\
92	&	174803.38-244641.4	&	267.01405	&	-24.77829	&	0.30	&	28	$\pm$	6	&	0.09	$\pm$	0.02	&	0.00	$\pm$	0.69	&	\nodata	\\
93	&	174806.10-244655.8	&	267.02535	&	-24.78229	&	0.30	&	27	$\pm$	6	&	0.09	$\pm$	0.02	&	-1.08	$\pm$	0.55	&	\nodata	\\
94	&	174804.89-244650.7	&	267.02034	&	-24.78080	&	0.32	&	25	$\pm$	8	&	0.08	$\pm$	0.03	&	-1.25	$\pm$	1.84	&	\nodata	\\
95	&	174804.88-244656.1	&	267.02030	&	-24.78236	&	0.32	&	25	$\pm$	6	&	0.08	$\pm$	0.02	&	-0.94	$\pm$	0.50	&	\nodata	\\
96	&	174803.89-244630.8	&	267.01616	&	-24.77532	&	0.32	&	24	$\pm$	6	&	0.12	$\pm$	0.03	&	0.37	$\pm$	0.82	&	\nodata	\\
97	&	174809.04-244545.2$^*$	&	267.03767	&	-24.76255	&	0.33	&	23	$\pm$	5	&	0.11	$\pm$	0.02	&	0.55	$\pm$	0.46	&	\nodata	\\
98	&	174808.36-244615.9	&	267.03473	&	-24.77124	&	0.33	&	23	$\pm$	5	&	0.11	$\pm$	0.02	&	1.06	$\pm$	0.61	&	p	\\
99	&	174805.81-244603.3	&	267.02417	&	-24.76765	&	0.34	&	22	$\pm$	5	&	0.11	$\pm$	0.02	&	2.50	$\pm$	3.81	&	\nodata	\\
100	&	174804.71-244621.4	&	267.01957	&	-24.77270	&	0.36	&	19	$\pm$	5	&	0.06	$\pm$	0.02	&	-0.59	$\pm$	1.00	&	\nodata	\\
101	&	174803.24-244628.0	&	267.01343	&	-24.77452	&	0.36	&	19	$\pm$	5	&	0.09	$\pm$	0.02	&	1.44	$\pm$	1.65	&	\nodata	\\
102	&	174804.77-244655.9	&	267.01985	&	-24.78233	&	0.37	&	18	$\pm$	6	&	0.06	$\pm$	0.02	&	0.00	$\pm$	0.97	&	\nodata	\\
103	&	174804.99-244556.0	&	267.02077	&	-24.76562	&	0.37	&	18	$\pm$	5	&	0.06	$\pm$	0.02	&	-2.26	$\pm$	3.27	&	\nodata	\\
104	&	174807.37-244641.2	&	267.03069	&	-24.77820	&	0.37	&	18	$\pm$	5	&	0.06	$\pm$	0.02	&	-0.49	$\pm$	1.01	&	\nodata	\\
105	&	174805.04-244619.7	&	267.02102	&	-24.77218	&	0.38	&	17	$\pm$	5	&	0.06	$\pm$	0.02	&	-3.01	$\pm$	7.60	&	\nodata	\\
106	&	174801.55-244617.5	&	267.00647	&	-24.77163	&	0.38	&	17	$\pm$	5	&	0.06	$\pm$	0.02	&	-0.66	$\pm$	1.16	&	\nodata	\\
107	&	174803.40-244635.4	&	267.01430	&	-24.77659	&	0.38	&	17	$\pm$	5	&	0.08	$\pm$	0.02	&	1.67	$\pm$	2.19	&	\nodata	\\
108	&	174807.52-244639.7	&	267.03135	&	-24.77779	&	0.38	&	17	$\pm$	5	&	0.08	$\pm$	0.02	&	0.39	$\pm$	0.99	&	\nodata	\\
109	&	174806.39-244631.9	&	267.02654	&	-24.77566	&	0.38	&	17	$\pm$	5	&	0.06	$\pm$	0.02	&	-0.39	$\pm$	1.03	&	\nodata	\\
110	&	174803.70-244552.8	&	267.01540	&	-24.76479	&	0.39	&	16	$\pm$	4	&	0.05	$\pm$	0.01	&	-0.55	$\pm$	1.13	&	\nodata	\\
111	&	174801.28-244555.0$^*$	&	267.00532	&	-24.76527	&	0.41	&	15	$\pm$	4	&	0.07	$\pm$	0.02	&	0.75	$\pm$	0.70	&	\nodata	\\
112	&	174802.32-244646.8$^*$	&	267.00965	&	-24.77968	&	0.41	&	15	$\pm$	4	&	0.05	$\pm$	0.01	&	-0.44	$\pm$	1.14	&	\nodata	\\
113	&	174800.37-244725.8$^*$	&	267.00156	&	-24.79051	&	0.44	&	13	$\pm$	4	&	0.04	$\pm$	0.01	&	-1.31	$\pm$	1.84	&	\nodata	\\
114	&	174807.16-244729.0	&	267.02980	&	-24.79149	&	0.44	&	13	$\pm$	4	&	0.04	$\pm$	0.01	&	-0.51	$\pm$	1.19	&	\nodata	\\
115	&	174808.15-244658.4	&	267.03393	&	-24.78298	&	0.44	&	13	$\pm$	4	&	0.06	$\pm$	0.02	&	1.31	$\pm$	1.84	&	\nodata	\\
116	&	174803.80-244659.6	&	267.01577	&	-24.78334	&	0.44	&	13	$\pm$	4	&	0.04	$\pm$	0.01	&	-0.61	$\pm$	0.71	&	\nodata	\\
117	&	174803.55-244616.2	&	267.01473	&	-24.77126	&	0.46	&	12	$\pm$	4	&	0.06	$\pm$	0.02	&	2.60	$\pm$	5.44	&	\nodata	\\
118	&	174801.45-244653.2$^*$	&	267.00606	&	-24.78144	&	0.48	&	11	$\pm$	4	&	0.05	$\pm$	0.02	&	$>$ 1.74			&	\nodata	\\
119	&	174805.00-244553.0$^*$	&	267.02084	&	-24.76472	&	0.48	&	11	$\pm$	4	&	0.04	$\pm$	0.01	&	-1.06	$\pm$	1.85	&	\nodata	\\
120	&	174803.13-244703.3	&	267.01297	&	-24.78443	&	0.50	&	10	$\pm$	4	&	0.05	$\pm$	0.02	&	0.44	$\pm$	1.46	&	\nodata	\\
121	&	174801.08-244548.9$^*$	&	267.00450	&	-24.76332	&	0.50	&	10	$\pm$	4	&	0.03	$\pm$	0.01	&	$<$ -1.74			&	\nodata	\\
122	&	174805.11-244610.6$^*$	&	267.02128	&	-24.76960	&	0.50	&	10	$\pm$	4	&	0.05	$\pm$	0.02	&	1.51	$\pm$	2.74	&	\nodata	\\
123	&	174805.59-244712.0	&	267.02342	&	-24.78670	&	0.50	&	10	$\pm$	3	&	0.05	$\pm$	0.01	&	$>$ 1.74			&	\nodata	\\
124	&	174804.47-244648.7	&	267.01859	&	-24.78029	&	0.52	&	9	$\pm$	4	&	0.03	$\pm$	0.01	&	$<$ -1.61			&	\nodata	\\
125	&	174805.50-244709.7$^*$	&	267.02293	&	-24.78603	&	0.56	&	8	$\pm$	3	&	0.03	$\pm$	0.01	&	0.00	$\pm$	1.21	&	\nodata	\\
126	&	174807.64-244625.9	&	267.03196	&	-24.77384	&	0.64	&	6	$\pm$	2	&	0.03	$\pm$	0.01	&	0.75	$\pm$	1.72	&	\nodata	\\
127	&	174805.28-244549.6	&	267.02194	&	-24.76387	&	0.78	&	4	$\pm$	2	&	0.02	$\pm$	0.01	&	$>$ 1.11			&	\nodata		 
\label{tab:XraySources} 
\end{longtable}
\end{document}